\documentclass[aps,prd,showpacs,preprint,english]{revtex4}
\usepackage[latin1]{inputenc}
\usepackage{babel}
\usepackage{graphicx}

\def\sc{\scriptsize}
\def\beq{\begin{equation}}
\def\eeq{\end{equation}}
\def\beqn{\begin{eqnarray}}
\def\eeqn{\end{eqnarray}}

\begin{document}

\preprint{JLAB-THY-03-199}
\preprint{WM-03-110}
\title{Higher moments of nucleon spin structure 
functions in heavy baryon chiral perturbation theory and in a
resonance model}

\author{Chung-Wen Kao}
\affiliation{Department of Physics and Astronomy, University of
Manchester, Manchester, M13 9PL UK 
\footnote{Present address : Department of Physics, North Carolina State University, 
Raleigh, NC 27695 - 8202, USA}}
\author{D. Drechsel, S. Kamalov} 
\affiliation{Institut f\"{u}r Kernphysik, Johannes
Gutenberg-Universit\"{a}t, D-55099 Mainz, Germany}

\author{M. Vanderhaeghen} 
\affiliation{Department of Physics, College of William and Mary,
Williamsburg, VA 23187, USA}
\affiliation{Jefferson Laboratory, Newport News, VA 23606, USA}
\date{\today}
\begin{abstract}
The third moment $d_2$ of the
twist-3 part of the nucleon spin structure function $g_2$ 
is generalized to arbitrary momentum transfer $Q^2$ and is evaluated 
in heavy baryon chiral perturbation theory
(HBChPT) up to order ${\mathcal{O}}(p^4)$ 
and in a unitary isobar model (MAID).  
We show how to link $d_2$ as well as higher moments of the 
nucleon spin structure functions $g_1$ and $g_2$ 
to nucleon spin polarizabilities. 
We compare our results with the most recent experimental data, and 
find a good description of these available data within the 
unitary isobar model.  
We proceed to extract the twist-4 matrix element $f_2$ which appears in the 
$1/Q^2$ suppressed term in the twist expansion of the 
spin structure function $g_1$ for proton and neutron. 
\pacs{11.55.Hx, 13.60.Hb, 14.20.Dh} 
\end{abstract}
\maketitle

\section{Introduction}

The forward scattering of spacelike virtual photons (with virtuality
$Q^2$) on the nucleon, allows one to study sum rules which relate nucleon
structure quantities to inclusive electroproduction cross sections.
At large $Q^2$, it yields the sum rules studied in deep-inelastic
scattering (DIS) experiments (see Ref.~\cite{FJ01} for a recent
review). The study of such sum rules 
as function of $Q^2$ from the real photon point to large
$Q^2$, opens up the perspective to interpolate between the non-perturbative 
and perturbative regimes of QCD, as one goes from low to high $Q^2$. 
\newline
\indent
The moment $d_2$ of the nucleon spin structure functions 
can be measured by scattering longitudinally polarized
electron beams off nucleon targets with transverse and longitudinal 
polarizations. Being a higher moment in the Bjorken variable $x$,
$d_2$ contains appreciable contributions from the resonance region.
It is therefore the aim of this letter to study the threshold and
resonance contributions to $d_2$ within the frameworks of heavy baryon
chiral perturbation theory (HBChPT) and of a unitary isobar model (MAID).
\newline
\indent
Since a measurement of $d_2$ requires also transverse polarization,
experimental information on this observable has become available only
recently at SLAC~\cite{Abe96,E155X} and JLab~\cite{Ama03}. Further experiments
are underway or proposed at JLab~[Mez03]. In particular, the SLAC
experiments \cite{E155X} yielded the values~: 
$d_2^p = 0.0032 \pm 0.0017$ and   
$d_2^n = 0.0079 \pm 0.0048$ 
at $Q^2=5$~GeV$^2$.
\newline
\indent
At large momentum transfers, a non-zero value for $d_2$ directly measures 
a twist-3 quark-gluon matrix element in the nucleon, which has been 
estimated within several model calculations as well as within lattice QCD. 
In this paper, we show that when generalizing the 
definition of $d_2$ to arbitrary momentum transfers, 
its physical interpretation can be expressed in terms of generalized 
(i.e. $Q^2$ dependent) spin polarizabilities of the nucleon. 
These generalized polarizabilities appear in the analysis of the 
forward virtual Compton scattering amplitude (for a review, 
see Ref.~\cite{DPV03}). Using the calculation of these spin 
polarizabilities within HBChPT~\cite{KSV02} 
(see also Refs.~\cite{BHM02,BHM03}),  
and within a unitary isobar model~\cite{MAID}, 
we evaluate $d_2$. In an analogous way, the third moment of the spin 
structure function $g_1$ can be expressed through these spin polarizabilities. 
These generalized definitions for $d_2$ and the third moment of $g_1$ 
constitute useful observables to interpolate between a hadronic description 
at low $Q^2$ and a partonic description, based on the Operator Product 
Expansion (OPE) at large $Q^2$. 
\newline
\indent
Besides the twist-3 matrix element $d_2$, the $1/Q^2$ suppressed term in the 
twist expansion of the first moment of $g_1$ also contains  
the twist-4 matrix element $f_2$. It has been suggested 
(see e.g. Ref.\cite{FJ01}) that the matrix elements $d_2$ and $f_2$ 
enter in the response of the color electric and 
magnetic fields to the polarization of the nucleon in its rest frame, which 
can be expressed in terms of gluon field polarizabilities. 
Therefore an extraction of $f_2$, besides $d_2$, can yield us 
interesting new nucleon structure information. 
Such an phenomenological extraction was peformed for the first time 
in Ref.~\cite{JiMel97} based on the SLAC data of Ref.~\cite{Abe96}. 
With the advent of the recent JLab data  
for the first moment of the proton and neutron spin 
structure function $g_1$ in the intermediate $Q^2$ range, 
we may now perform a phenomenological analysis 
based on all available data to extract the twist-4 matrix element $f_2$. 
\newline
\indent
The outline of the paper is as follows. 
In Section~\ref{sec2}, we review the twist expansion of the moments  
of the nucleon spin structure functions $g_1$ and $g_2$ 
and show how the definitions 
for the third moments of the nucleon structure functions $g_1$ and $g_2$ 
can be generalized at arbitrary $Q^2$ in terms of 
nucleon spin polarizabilities. 
\newline
\indent
In Section~\ref{sec3}, we use the calculation of these spin 
polarizabilities within HBChPT 
to evaluate $d_2$ and the third moments of $g_1$ and $g_2$. 
\newline
\indent
In Section~\ref{sec4}, we show how these observables can be expressed in 
multipoles which arise naturally when performing an evaluation within an 
isobar model. 
\newline
\indent
In Section~\ref{sec5}, we show our results for $d_2$ and the third moments  
of $g_1$ and $g_2$ 
within both ChPT and a unitary isobar model and compare with  
recent experimental data for $d_2$. 
We also compare our results for the first moment of $g_1$ 
with the available data for both proton and neutron and use the 
twist-expansion for the first moment of $g_1$ to extract the magnitude of the 
twist-4 matrix element $f_2$. 
\newline
\indent
Finally, we give our conclusions in Section~\ref{sec6}.

\section{Twist expansion and moments of nucleon spin structure functions}
\label{sec2}

At sufficiently large momentum transfer $Q^2 >> \Lambda_{QCD}^2$, one can
perform a twist expansion for the moments of the nucleon structure functions. 
We will consider in this work the moments of the 
nucleon spin structure functions $g_1$ and $g_2$, which are defined as~:
\begin{eqnarray}
\Gamma_1^{(n)}(Q^2) \,&\equiv& \, 
\int_{0}^{1}\,dx \, x^{n - 1} \, g_1\,(x,\,Q^2) \, , 
\hspace{2.5cm} n = 1, 3, 5, ...  
\label{eq:momg1} \\
\Gamma_2^{(n)}(Q^2) \,&\equiv& \, 
\int_{0}^{1}\,dx \, x^{n - 1} \, g_2\,(x,\,Q^2) \, , 
\hspace{2.5cm} n = 1, 3, 5, ...  
\label{eq:momg2}
\end{eqnarray}
In particular, for the first moment $\Gamma_1^{(1)}$ of $g_1$  
such a twist expansion can be written as~\cite{Ji93,JiUn94}~: 
\begin{eqnarray}
\Gamma_1^{(1)}(Q^2) 
\, &=& \, \Gamma_{1, \, tw-2}^{(1)}(Q^2) \,+\, \frac{M_N^2}{9 \, Q^2} \,
\left( a_2(Q^2) \,+\, 4 d_2(Q^2) \,+\, 4 f_2(Q^2) \right) \,+\, 
{\mathcal{O}}(\frac{M_N^4}{Q^4}) \, .
\label{eq:g1twist}
\end{eqnarray}
The remaining $Q^2$ dependence in the coefficients 
$\Gamma_{1, \, tw-2}^{(1)}$, $a_2$, $d_2$ and $f_2$ of the twist expansion is
logarithmic. In particular, the three-loop result 
for the leading (i.e. twist-2) term $\Gamma_{1, \, tw-2}^{(1)}(Q^2)$  
is given, for 3 quark flavors, as~\cite{LV91,Lar94,LRV97}~:
\begin{eqnarray}
&&\Gamma_{1, \, tw-2}^{(1)}(Q^2) \nonumber \\
&&\hspace{0.2cm}\,=\, \left[ 1 - \left( \frac{\alpha_s(Q^2)}{\pi} \right)
- 3.5833\left( \frac{\alpha_s(Q^2)}{\pi} \right)^2
- 20.2153\left( \frac{\alpha_s(Q^2)}{\pi} \right)^3 \right] \,
\left( \pm \frac{1}{12} g_A \, \,+\, \frac{1}{36} a_8 \right) \nonumber \\
&&\hspace{0.2cm} 
\,+\, \left[ 1 - 0.33333 \left( \frac{\alpha_s(Q^2)}{\pi} \right)
- 0.54959 \left( \frac{\alpha_s(Q^2)}{\pi} \right)^2
- 4.44725 \left( \frac{\alpha_s(Q^2)}{\pi} \right)^3  
\right] \, \frac{1}{9} \, a_0^{\infty} \,  , 
\label{eq:mu2}
\end{eqnarray}
where in the term proportional to $g_A$ the sign + (-) 
corresponds with proton (neutron) respectively, 
$g_A$ is known from neutron beta decay, and 
$a_8$ is extracted from hyperon weak decay data assuming $SU(3)$
symmetry. The updated values for $g_A$~\cite{Hag02} 
and $a_8$~\cite{Got00}~ are given by~:
\begin{eqnarray}
g_A \,&=&\, 1.267 \pm 0.003 \, , \\
a_8 \,&=&\, 0.585 \pm 0.023 \, .
\end{eqnarray}
Furthermore, 
in Eq.~(\ref{eq:mu2}), $a_0^{\infty} \equiv a_0(Q^2 = \infty)$ is the
flavor singlet axial charge, which is fixed by measuring the 
first moment $\Gamma_1^{(1)}$ at a sufficiently large scale.  
\newline
\indent
The $1/Q^2$ suppressed terms in Eq.~(\ref{eq:g1twist}) are of three
different natures. 
The term proportional to $a_2(Q^2)$ arises due to target mass
corrections and is given by the twist-2 part of the third moment of
$g_1$~:
\begin{eqnarray}
a_2(Q^2) \,&\equiv& \, 2 \, \Gamma^{(3)}_{1, \, tw-2}(Q^2) \, .
\label{eq:a2}
\end{eqnarray}
\newline
\indent
The terms proportional to $d_2$ and $f_2$ in Eq.~(\ref{eq:g1twist}) 
correspond with dynamical higher-twist corrections. 
The function $d_2$ corresponds with the matrix element of a
twist-3 quark gluon operator, and can be expressed in terms of the
twist-3 part of the spin structure function $g_2$ as~:
\begin{eqnarray}
d_2(Q^2) 
&\,\equiv\,& 3 \, \int_{0}^{1}\,dx \, x^2 \, 
\left(g_2\,(x,\,Q^2) - g_2^{WW}(x,\,Q^2) \right)\, ,
\label{eq:d2def}
\end{eqnarray}
where $g_2^{WW}$ is the twist-2 (Wandzura-Wilczek) part of $g_2$~\cite{WW77}. 
Using the Wandzura-Wilczek relation, one can express
Eq.~(\ref{eq:d2def}) as~:
\begin{eqnarray}
d_2(Q^2) 
&\,=\,& \int_{0}^{1}\,dx \, x^2 \, 
\left(3 \, g_2\,(x,\,Q^2) + 2 \, g_1(x,\,Q^2) \right)\, .
\label{eq:d2}
\end{eqnarray}
Several model estimates  
as well as lattice QCD calculations~\cite{Gock01} 
have been performed for the twist-3 matrix element $d_2$.  
In particular, an estimate in the instanton vacuum
approach~\cite{Lee02}, where $d_2$ is parametrically suppressed due to the
diluteness of the instanton medium, predicted $d_2$ to be of order
$10^{-3}$. Also a revised lattice calculation \cite{Gock01} supports  
this small value. 
Such a small value for $d_2$, of order $10^{-3}$,  
is in agreement with the recent experimental results from SLAC \cite{E155X}. 
\newline
\indent
The function $f_2$ in Eq.~(\ref{eq:g1twist}) 
corresponds with the matrix element of a twist-4 quark gluon
operator. 
Unlike $a_2$ and $d_2$, 
the matrix element $f_2$ cannot be expressed directly in terms of 
moments of $g_1$ and $g_2$ as in Eqs.~(\ref{eq:a2}) and (\ref{eq:d2}). 
It can however be extracted phenomenologically 
from the twist expansion of Eq.~(\ref{eq:g1twist}) as~:
\begin{eqnarray}
\frac{4 M_N^2}{9 Q^2} \, f_2(Q^2) \,\equiv\,   
\left[ \Gamma^{(1)}_1(Q^2) - \Gamma^{(1)}_{1, \, tw-2}(Q^2) \right] \,-\, 
\frac{M_N^2}{9 Q^2} \left[ a_2(Q^2) + 4 d_2(Q^2) \right] \, ,
\label{eq:f2def}
\end{eqnarray}
by using the experimental information on the full $Q^2$ dependence of 
$\Gamma^{(1)}_1$ 
and provided one knows the twist-2 and twist-3 
contributions $\Gamma^{(1)}_{1, \, tw-2}$, $a_2$ and $d_2$. 
In this way, $f_2$ has been estimated in Refs.~\cite{JiMel97,Ede00} 
for $Q^2 \gtrsim 0.5$ GeV$^2$ where the twist expansion was assumed
to hold. 
Several model calculations have also been given for $f_2$, e.g. 
in the framework of the instanton vacuum approach \cite{Lee02} 
and using QCD sum rules \cite{Ste95}, to which we refer further on.  
\newline
\indent
Although the Operator Product Expansion (OPE) 
of Eq.~(\ref{eq:g1twist}) is only defined for $Q^2 >>
\Lambda_{QCD}^2$, Eq.~(\ref{eq:d2}) 
can be used to define $d_2$ outside this range
by keeping the full $Q^2$ dependence in the structure functions which
appear in the integrand. 
Analogously, we can study the $Q^2$ dependence of the higher moments
of $g_1$, such as $\Gamma^{(3)}_1(Q^2)$, outside the range of validity
of the OPE.  
We next show that at low $Q^2$, the physical meaning of $d_2$ and
$\Gamma^{(3)}_1$ can be expressed in terms of nucleon polarizabilities 
which will be calculated in this work within 
Chiral Perturbation Theory and estimated within a phenomenological
resonance approach. At large $Q^2$, $d_2$ and $\Gamma^{(3)}_1$ 
tend into the matrix elements
appearing in the twist expansion of Eq.~(\ref{eq:g1twist}) and display 
only a logarithmic $Q^2$ dependence. Defined in this way, 
$d_2$ and $\Gamma^{(3)}_1$ are useful observables to interpolate between 
a hadronic description at low $Q^2$, 
involving the polarizabilities of the system,  
and a partonic description based on the OPE at large $Q^2$.     
\newline
\indent
Starting with $d_2$, one can split the integral of Eq.~(\ref{eq:d2}) 
into an elastic contribution at $x = 1$
and an inelastic contribution. 
The elastic contribution is given by~:
\begin{eqnarray}
d_2^{el}(Q^2) 
&\,=\,& \left( G_E(Q^2) \,+\, 
\frac{G_E(Q^2) - G_M(Q^2)}{2 (1 + 4 M_N^2 / Q^2)}\right)\, G_M(Q^2).
\label{eq:d2el}
\end{eqnarray}
While this contribution vanishes like $Q^{-8}$ for $Q^2 \to \infty$,
it increases with smaller values of $Q^2$ and approaches 
$d_2^{\mathrm{el}}(0) = G_E(0) \,\cdot \, G_M(0)$. 
The inelastic contribution to $d_2$ corresponds with the integral over
the excitation spectrum~:
\begin{eqnarray}
d_2^{inel}(Q^2) 
&\,=\,& \int_{0}^{x_0}\,dx \, x^2 \, 
\left(3 \, g_2\,(x,\,Q^2) + 2 \, g_1(x,\,Q^2) \right)\, .
\label{eq:d2inel}
\end{eqnarray}
where $x_0 = Q^2 / (2 M_N m_\pi + m_\pi^2 + Q^2)$ is the threshold for
pion production.
\newline
\indent
To evaluate Eq.~(\ref{eq:d2inel}), we can 
equivalently express the third moment of the twist-3 part of $g_2$ in
terms of the spin-dependent doubly virtual Compton scattering
amplitude in the forward direction (VVCS). Following the notations 
of Refs.~\cite{DPV03,KSV02},
we use the VVCS amplitudes 
$g_{TT}(\nu,Q^2)$ and $g_{LT}(\nu,Q^2)$, 
where $\nu$ is the lab energy and $Q^2$ the virtuality 
of the virtual photon,
and T (L) denotes the transverse
(longitudinal) virtual photon polarization. 
\newline
\indent
The imaginary parts of $g_{TT}$ and $g_{LT}$ are related to the 
virtual photon absorption cross sections 
$\sigma_{TT}$ and $\sigma_{LT}$, multiplied by a photon flux factor
$K$ (with dimension of energy)
\footnote{Note that the partial cross sections $\sigma_{TT}$ and 
$\sigma_{LT}$ depend on the virtual photon flux convention. However, 
in the dispersion integrals only 
the products $K \cdot \sigma_{TT}$ and $K \cdot \sigma_{LT}$ enter,  
which are independent of this convention.}.
These partial cross sections are related to the nucleon structure
functions $g_1$ and $g_2$ as~:
\begin{eqnarray}
K \cdot \sigma_{TT} \,&=&\, \frac{4\pi^{2}\alpha_{em}}{M_N}  
\biggl( g_{1}(x, Q^2) -\gamma^{2}g_{2}(x, Q^2) \biggr), 
\label{eq:stt} \\
K \cdot \sigma_{LT} \,&=&\, \frac{4\pi^{2}\alpha_{em}}{M_N} 
\gamma \biggl( g_{1}(x, Q^2) + g_{2}(x, Q^2) \biggr) \, ,
\label{eq:slt}
\end{eqnarray}
with $\gamma \equiv Q/\nu$ and $x \equiv Q^2 / (2 M_N \, \nu)$.
\newline
\indent
For the non-pole (i.e. {\it inelastic}) 
contributions to $g_{TT}$ and $g_{LT}$, one can
perform a low energy expansion (LEX) as follows \cite{DPV03}~:
\begin{eqnarray}
{\mbox{Re}}\ g_{TT}(\nu,\,Q^2) \,-\,
{\mbox{Re}}\ g_{TT}^{\mbox{\sc{pole}}}(\nu,\,Q^2) &=&
\left(\frac{2 \, \alpha_{\mbox{\sc{em}}} }{M_N^2} \right) \, I_A(Q^2) \; \nu
\,+\, \gamma_0(Q^2) \; \nu^3 \,+\, {\mathcal{O}}(\nu^5) \, , 
\label{eq:gttlex} \\
{\mbox{Re}}\ g_{LT}(\nu,\,Q^2) -
{\mbox{Re}}\ g_{LT}^{\mbox{\sc{pole}}}(\nu,\,Q^2) &=&
\left(\frac{2 \, \alpha_{\mbox{\sc{em}}} }{M_N^2} \right) Q \, I_3(Q^2)
\,+\, Q \, \delta_{LT}(Q^2) \, \nu^2 \,+\, {\mathcal{O}}(\nu^4) \, .
\label{eq:gltlex}
\end{eqnarray}
For the ${\mathcal{O}}(\nu)$ term in Eq.~(\ref{eq:gttlex}), one
obtains a generalization of the GDH sum rule as~:
\begin{eqnarray}
\label{eq:ia}
I_A(Q^2) &\,=\,&
\frac{M_N^2}{4 \, \pi^2 \, \alpha_{\mbox{\sc{em}}}}\,
\int_{\nu_0}^{\infty}\, \frac{K(\nu, \, Q^2)}{\nu} \,
\frac{\sigma_{TT}\,(\nu,\,Q^2)}{\nu}\,d\nu \, , \nonumber\\
&\,=\,& \frac{2 \, M_N^2}{Q^2}\,
\int_{0}^{x_0}\,dx \, \left\{ g_1\,(x,\,Q^2)
\,-\, \frac{4 M_N^2}{Q^2} \, x^2 \, g_2\,(x,\,Q^2) \right\} \, ,
\end{eqnarray}
and recovers the GDH sum rule at $Q^2$ = 0, as  $I_A(0) = - \kappa_N^2
/ 4$, with $\kappa_N$ the nucleon anomalous magnetic moment ($\kappa_p
= 1.79$, $\kappa_n = -1.91$).
\newline
\indent
Furthermore we introduce the integral $I_1$, which is related to the 
inelastic part of the first moment $\Gamma_1^{(1)}$ of $g_1$ as : 
\begin{eqnarray}
\label{eq:i1}
I_1(Q^2) 
&\,=\,& \frac{2 \, M_N^2}{Q^2}\,
\int_{0}^{x_0}\,dx \, g_1\,(x,\,Q^2) \, ,
\end{eqnarray}
and which also approaches the GDH value at $Q^2 = 0$. 
\newline
\indent
The higher order terms in Eqs.~(\ref{eq:gttlex}) and (\ref{eq:gltlex}) 
can be expressed in terms of nucleon spin polarizabilities, see 
Ref.~\cite{DPV03}.
In particular, the ${\mathcal{O}}(\nu^2)$ term in $g_{LT}$ is given by~:
\begin{eqnarray}
\label{eq:deltalt}
\delta_{LT}\,(Q^2) \,&=&\, \frac{1}{2\pi^2}\,
\int_{\nu_0}^{\infty}\,\frac{K(\nu, \, Q^2)}{\nu} \,
\frac{\sigma_{LT}(\nu\,Q^2)}{Q\,\nu^2}\,d\nu  \, , \nonumber \\
\,&=&\, \frac{\alpha_{\mbox{\sc{em}}} \, 16 \, M_N^2}{Q^6}\,
\int_{0}^{x_0}\,dx \, x^2 \,
\left\{ g_1\,(x,\,Q^2) \,+\, g_2\,(x,\,Q^2) \right\} \, .
\end{eqnarray}
\indent
Combining Eqs.~(\ref{eq:ia},\ref{eq:i1},\ref{eq:deltalt}), we find
that the inelastic contribution to the 
third moment $d_2$ of Eq.~(\ref{eq:d2inel}) can be expressed as~:
\begin{eqnarray}
d_2^{inel}(Q^2) \,=\, {{Q^6} \over {8 M_N^2}}
\left\{- {1 \over {M_N^2 \, Q^2}} \left( I_A(Q^2) - I_1(Q^2) \right) 
\,+\, \frac{1}{\alpha_{\mbox{\sc{em}}}} \,
\delta_{LT}(Q^2) \right\} \, .
\label{eq:d2b}
\end{eqnarray}
\indent
We can perform a similar analysis for the moments
$\Gamma^{(n)}_1$ and $\Gamma^{(n)}_2$ 
defined through Eqs.~(\ref{eq:momg1}) and (\ref{eq:momg2}).
The elastic contribution to $\Gamma^{(n)}_1$ is given by~:
\begin{eqnarray}
\Gamma_1^{(n) \, el}(Q^2) 
&\,=\,& \frac{1}{2} \, \frac{1}{1 + Q^2 / (4 M_N^2)} \, 
\left( G_E(Q^2) \,+\, \frac{Q^2}{4 M_N^2} \, G_M(Q^2) \right)\, G_M(Q^2).
\label{eq:momg1el}
\end{eqnarray}
While this contribution vanishes like $Q^{-8}$ for $Q^2 \to \infty$,
it approaches 
$\Gamma_1^{(n) \, el}(0) = 1/2 \, \cdot \, G_E(0) \,\cdot \, G_M(0)$ at the
real photon point. 
The inelastic contribution to $\Gamma_1^{(n)}$ 
corresponds with the integral over the excitation spectrum~:
\begin{eqnarray}
\Gamma_1^{(n) \, inel}(Q^2) 
&\,=\,& \int_{0}^{x_0}\,dx \, x^{n - 1} \, g_1(x,\,Q^2) \, .
\label{eq:momg1inel}
\end{eqnarray}
The third moment $\Gamma_1^{(3) \, inel}$ can be expressed in terms of
the quantities introduced in 
Eqs.~(\ref{eq:ia},\ref{eq:i1},\ref{eq:deltalt}) as~:
\begin{eqnarray}
\Gamma_1^{(3) \, inel}(Q^2) \,=\, {{Q^6} \over {8 M_N^2}}
\left\{ {1 \over {M_N^2 \, Q^2}} \left( I_A(Q^2) - I_1(Q^2) \right) 
\,+\, \frac{1}{2 \, \alpha_{\mbox{\sc{em}}}} \, 
\delta_{LT}(Q^2) \right\} \, .
\label{eq:mom3g1inel}
\end{eqnarray}
\indent
For $\Gamma^{(n)}_2$, the elastic contribution is given by~:
\begin{eqnarray}
\Gamma_2^{(n) \, el}(Q^2) 
&\,=\,& -\frac{1}{2} \, \frac{Q^2}{4 M_N^2 + Q^2} \, 
\left( G_M(Q^2) \,-\, G_E(Q^2) \right)\, G_M(Q^2).
\label{eq:momg2el}
\end{eqnarray}
This contribution vanishes at the real photon point. 
The inelastic contribution to $\Gamma_2^{(n)}$ 
corresponds with the integral over the excitation spectrum~:
\begin{eqnarray}
\Gamma_2^{(n) \, inel}(Q^2) 
&\,=\,& \int_{0}^{x_0}\,dx \, x^{n - 1} \, g_2(x,\,Q^2) \, .
\label{eq:momg2inel}
\end{eqnarray}
The third moment $\Gamma_2^{(3) \, inel}$ can be expressed in terms of
the quantities introduced in Eqs.~(\ref{eq:ia},\ref{eq:i1}) as~:
\begin{eqnarray}
\Gamma_2^{(3) \, inel}(Q^2) \,=\, -{{Q^4} \over {8 M_N^4}}
\left( I_A(Q^2) - I_1(Q^2) \right) \, .
\label{eq:mom3g2inel}
\end{eqnarray}
Note that  $\Gamma_2^{(3)}$ is linearly dependent on $d_2$ and 
$\Gamma_1^{(3)}$ as it can be expressed as~:
\begin{eqnarray}
\Gamma_2^{(3)}(Q^2) \,=\, {1 \over 3}
\left( d_2(Q^2) - 2 \, \Gamma_1^{(3)}(Q^2) \right) \, .
\label{eq:mom3g2equiv}
\end{eqnarray}
\indent
Although we restrict our investigation in this work 
up to the third moment of $g_1$ and $g_2$, 
one can in principle extend the analysis 
to the higher moments $\Gamma_1^{(n)}$ and 
$\Gamma_2^{(n)}$, with $n = 5, 7 ...$. Their
expressions involve higher spin polarizabilities, as have e.g. been
introduced at the real photon point in Ref.~\cite{Hol00}.

\section{Calculation of the moments ${\lowercase{d}}_2$, 
$\Gamma_1^{(3)}$, and $\Gamma_2^{(3)}$ in HBC\lowercase{h}PT}
\label{sec3}

Eqs.~(\ref{eq:d2b},\ref{eq:mom3g1inel},\ref{eq:mom3g2inel}) 
show that $d_2^{inel}(Q^2)$, $\Gamma_1^{(3) \, inel}(Q^2)$ and 
$\Gamma_2^{(3) \, inel}(Q^2)$ can be expressed in terms of $I_A$, $I_1$ and
$\delta_{LT}$ appearing in a low energy expansion of the forward 
VVCS amplitudes . These expressions have been calculated at low $Q^2$ in
HBChPT, which allow us to construct the HBChPT predictions 
for $d_2^{inel}$, $\Gamma_1^{(3) \, inel}$ and $\Gamma_2^{(3) \, inel}$.
The expressions for $I_A$ and $I_1$ in HBChPT 
have been obtained up to ${\mathcal{O}}(p^4)$ in Refs.~\cite{Ji01,JKO00}~:
\begin{eqnarray}
I_{1}(Q^2)
&=&- \frac{1}{16}[(\kappa_{s} + \kappa_{v}\tau_{3})^2] \nonumber \\
&+&\frac{g_{A}^2 M_{N}^2}{(4\pi F_{\pi})^2}\cdot {{m_\pi} \over {M_N}}
\cdot\frac{\pi}{32}
\{(-10-12\kappa_{v})+(-2-12\kappa_{s})\tau_{3} \nonumber \\
&&\hspace{3.5cm}+\,[(20+24\kappa_{v})+(4+24\kappa_{s})\tau_{3}]
\cdot\frac{1}{w}\tan^{-1}[\frac{w}{2}] \nonumber \\ 
&&\hspace{3.5cm}+\,[(3+6\kappa_{v})+(3+10\kappa_{s})\tau_{3}]\cdot
w\tan^{-1}[\frac{w}{2}]\} \, , 
\label{eq:i1chpt} \\
I_{A}(Q^2)
&=&- \frac{1}{16}[(\kappa_{s} + \kappa_{v}\tau_{3})^2] \nonumber \\
&+& \frac{g_{A}^2 M_{N}^2}{(4\pi F_{\pi})^2}\, 2 \,
\left({{\sqrt{w^2 + 4}} \over {w}} \, \sinh^{-1} [{w \over 2}] 
\,-\, 1\right) \nonumber \\
&-&\frac{g_{A}^2 M_{N}^2}{(4\pi F_{\pi})^2}\cdot{{m_{\pi}} \over {M_N}}
\cdot\frac{\pi}{16}
\{(-10-2\kappa_{v}+8 w^2)+(-2+6 \kappa_{s})\tau_{3} \nonumber \\
&&\hspace{3.5cm}+\,[(20+4\kappa_{v})+(4-12\kappa_{s})\tau_{3}]
\cdot\frac{1}{w}\tan^{-1}[\frac{w}{2}] \nonumber \\ 
&&\hspace{3.5cm}+\,[(3+3\kappa_{v})+(3-\kappa_{s})\tau_{3}]\cdot
w\tan^{-1}[\frac{w}{2}]\} \, , 
\label{eq:iachpt}
\end{eqnarray}
with $w=\sqrt{\frac{Q^2}{m_{\pi}^{2}}}$.
These expressions are given  
in terms of the renormalized isoscalar ($\kappa_s$) and isovector 
($\kappa_v$) anomalous magnetic moments, 
whose physical values are given by 
$\kappa_s = -0.12$ and $\kappa_v = 3.70$.
Furthermore, throughout this paper we use the values 
$g_A = 1.267$, $F_\pi = 0.0924$~GeV, and $m_\pi = 0.14$~GeV. 
While both $I_1$ of Eq.~(\ref{eq:i1chpt}) and $I_A$ of
Eq.~(\ref{eq:iachpt}) approach the GDH value of $- \kappa_N^2 / $ at
$Q^2 = 0$, their slopes at this point differe substantially~:
$I_1^{\, '}(0)$ = (6.31 + 0.66 $\tau_3$) / GeV$^2$ and 
$I_A^{\, '}(0)$ = - (12.43 + 2.09 $\tau_3$) / GeV$^2$.
\newline
\indent
The longitudinal-transverse generalized
forward spin polarizability $\delta_{LT}$ of Eq.~(\ref{eq:gltlex}) 
has also been calculated in HBChPT~\cite{KSV02} (see Ref.~\cite{BHM03} for the 
corresponding calculation within the framework of relativistic baryon ChPT).
For the  ${\mathcal{O}}(p^3)$ term, the result is~:
\begin{eqnarray}
\delta_{LT}^{{\cal O}(p^3)}(Q^2)=\frac{\alpha_{em}g_{A}^{2}}{(4\pi F_{\pi})^{2}}\cdot\frac{4}{m_{\pi}^{2}}\left[
\frac{1}{3w\sqrt{w^2+4}}\sinh^{-1}[\frac{w}{2}]\right] \, ,
\label{eq:dltp3}
\end{eqnarray}
and the ${\mathcal{O}}(p^4)$ correction is given by~:
\begin{eqnarray}
\delta_{LT}^{{\cal O}(p^4)}(Q^2)=
\frac{\alpha_{em}g_{A}^2}{(4\pi F_{\pi})^{2} m_{\pi}^2}
\cdot \frac{m_\pi}{M_N} \cdot\frac{\pi}{64}
&&\{(-16+8\kappa_{v})+(-8+16\kappa_{s})\tau_{3} \nonumber \\
&&+[(-54+8\kappa_{v})+(-6+8\kappa_{s})\tau_{3}]\cdot\frac{1}{w^2}
\nonumber \\
&&+[(-9-12\kappa_{v})+(-9-4\kappa_{s})\tau_{3}]\cdot
\frac{1}{w}\tan^{-1}[\frac{w}{2}] \nonumber \\
&&+[(-84-16\kappa_{v})+(12-16\kappa_{s})\tau_{3}]
\cdot\frac{1}{w^3}\tan^{-1}[\frac{w}{2}] \nonumber \\
&&+[4-(12+16\kappa_{s})\tau_{3}]\cdot\frac{1}{4+w^2}
+128\cdot\frac{3+w^2}{4w^2+w^4} \} \, .
\label{eq:dltp4}
\end{eqnarray}
\newline
\indent
Using
Eqs.~(\ref{eq:i1chpt} - \ref{eq:dltp4}),
we can now construct the result of $d_2^{inel}$ from Eq.~(\ref{eq:d2b}).
The result at ${\mathcal{O}}(p^3)$ for $d_2^{inel}$ is
\begin{eqnarray}
d_2^{{\cal O}(p^3)}(Q^2) &=& \frac{Q^6}{8 M_N^2} 
\frac{g_{A}^2}{(4\pi F_{\pi})^{2} m_{\pi}^2}
\left({{-24 - 2 w^2} \over {3 w^3 \sqrt{w^2 + 4}}} \, \sinh^{-1} [{w \over 2}] 
\,+\, {{2} \over {w^2}}\right) \, , 
\label{eq:d2p3}
\end{eqnarray}
and the ${\mathcal{O}}(p^4)$ correction to $d_2^{inel}$ is given by~:
\begin{eqnarray}
d_2^{{\cal O}(p^4)}(Q^2) &=& \frac{Q^6}{8 M_N^2} 
\frac{g_{A}^2}{(4\pi F_{\pi})^{2} m_{\pi}^2}
\cdot \frac{m_\pi}{M_N} \cdot\frac{\pi}{32} \nonumber\\
&\times& \left\{
\left[(-8+4\kappa_{v})+(-4+8\kappa_{s})\tau_{3} \right] \right. \nonumber \\
&&+\left[(-57-12\kappa_{v}+16 w^2)+(-9+4\kappa_{s})\tau_{3} \right]
\cdot\frac{1}{w^2} \nonumber \\
&&+\frac{1}{2} \left[(9+12\kappa_{v})+(9+12\kappa_{s})\tau_{3} \right]\cdot
\frac{1}{w}\tan^{-1}[\frac{w}{2}] \nonumber \\
&&+\left[(18+24\kappa_{v})+(18-8\kappa_{s})\tau_{3} \right]
\cdot\frac{1}{w^3}\tan^{-1}[\frac{w}{2}] \nonumber \\
&&\left. 
+\left[2 + 64 \cdot \frac{3 + w^2}{w^2} -(6+8\kappa_{s})\tau_{3} \right]
\cdot\frac{1}{4+w^2} \right\} \, .
\label{eq:d2p4}
\end{eqnarray}
\indent
Analogously, we can also construct the results for 
$\Gamma_1^{(3) \, inel}$ and $\Gamma_2^{(3) \, inel}$  
of Eqs.~(\ref{eq:mom3g1inel},\ref{eq:mom3g2inel}) 
in HBChPT by using the expressions of 
Eqs.~(\ref{eq:i1chpt} - \ref{eq:dltp4}). 
The result at ${\mathcal{O}}(p^3)$ are given by~:
\begin{eqnarray}
\Gamma_1^{(3)\, {\cal O}(p^3)}(Q^2) &=&\frac{Q^6}{4 M_N^2} 
\frac{g_{A}^2}{(4\pi F_{\pi})^{2} m_{\pi}^2}
\left({{12 + 4 w^2} \over {3 w^3 \sqrt{w^2 + 4}}} \, \sinh^{-1} [{w \over 2}] 
\,-\, {{1} \over {w^2}}\right) \, , 
\label{eq:a2p3} \\
\Gamma_2^{(3)\, {\cal O}(p^3)}(Q^2) &=&- \frac{Q^6}{4 M_N^2} 
\frac{g_{A}^2}{(4\pi F_{\pi})^{2} m_{\pi}^2}
\left({{4 + w^2} \over {w^3 \sqrt{w^2 + 4}}} \, \sinh^{-1} [{w \over 2}] 
\,-\, {{1} \over {w^2}}\right) \, , 
\label{eq:g23_chpt3}
\end{eqnarray}
and the ${\mathcal{O}}(p^4)$ corrections are given by~:
\begin{eqnarray}
\Gamma_1^{(3)\, {\cal O}(p^4)}(Q^2) &=& \frac{Q^6}{8 M_N^2} 
\frac{g_{A}^2}{(4\pi F_{\pi})^{2} m_{\pi}^2}
\cdot \frac{m_\pi}{M_N} \cdot\frac{\pi}{64} \nonumber\\
&\times& \left\{
\left[(-40+4\kappa_{v})+(-4+8\kappa_{s})\tau_{3} \right] \right. \nonumber \\
&&+\left[(33+36\kappa_{v})+(9+4\kappa_{s})\tau_{3} \right]
\cdot\frac{1}{w^2} \nonumber \\
&&+\frac{1}{2} \left[(-45-60\kappa_{v})+(-45-36\kappa_{s})\tau_{3} \right]\cdot
\frac{1}{w}\tan^{-1}[\frac{w}{2}] \nonumber \\
&&+\left[(-162-72\kappa_{v})+(-18-8\kappa_{s})\tau_{3} \right]
\cdot\frac{1}{w^3}\tan^{-1}[\frac{w}{2}] \nonumber \\
&&\left. 
+\left[2 + 64 \cdot \frac{3 + w^2}{w^2} -(6+8\kappa_{s})\tau_{3} \right]
\cdot\frac{1}{4+w^2} \right\} \, , 
\label{eq:a2p4} \\
\Gamma_2^{(3)\, {\cal O}(p^4)}(Q^2) &=& \frac{Q^6}{8 M_N^2} 
\frac{g_{A}^2}{(4\pi F_{\pi})^{2} m_{\pi}^2}
\cdot \frac{m_\pi}{M_N} \cdot\frac{\pi}{32}  \nonumber\\
&\times& \left\{
\left[(-30-16\kappa_{v} + 16 w^2) - 6 \tau_{3} \right] 
\cdot \frac{1}{w^2} \right. \nonumber \\
&&+ \left[(9 + 12\kappa_{v}) + (9 + 8\kappa_{s})\tau_{3} \right] \cdot
\frac{1}{w}\tan^{-1}[\frac{w}{2}] \nonumber \\
&&\left. +
\left[(60 + 32 \kappa_{v}) + 12 \tau_{3} \right]
\cdot\frac{1}{w^3}\tan^{-1}[\frac{w}{2}] \right\} \, .
\label{eq:g23_chpt4}
\end{eqnarray}
From Eqs.~(\ref{eq:d2p3} - \ref{eq:g23_chpt4}), one can extract that the 
following prediction of HBChPT in the limit $Q^2 \to 0$~:
\begin{eqnarray}
d_2^{inel}(Q^2) &\longrightarrow& 
\frac{Q^6}{8 M_N^2} \, 
\frac{1}{2 \alpha_{em}} \, 
\left( - \gamma_0(0) + 3 \delta_{LT}(0) \right)  \nonumber \\
&=&\, \frac{Q^6}{48 M_N^2} 
\frac{g_{A}^2}{(4\pi F_{\pi})^{2} \, m_{\pi}^2}
\left\{ 1 \,+\, {{m_\pi} \over {M_N}} \cdot {\pi \over 8} 
\left[ (21 + 9 \kappa_v) + (-6 + 14 \kappa_s) \tau_3 \right] \right\} \, .
\label{eq:d2limit} \\
\Gamma_1^{(3) \, inel}(Q^2) &\longrightarrow&  \frac{Q^6}{8 M_N^2} \, 
\frac{1}{2 \, \alpha_{em}} \, \gamma_0(0) \nonumber \\
&=&\,\frac{Q^6}{48 M_N^2} \frac{g_{A}^2}{(4\pi F_{\pi})^{2} \, m_{\pi}^2}
\left\{ 2 \,+\, {{m_\pi} \over {M_N}} \cdot {\pi \over 4} 
\left[ (-15 - 3 \kappa_v) + (-6 - \kappa_s) \tau_3 \right] \right\} \, ,
\label{eq:a2limit} \\
\Gamma_2^{(3) \, inel}(Q^2) &\longrightarrow&  \frac{Q^6}{8 M_N^2} \, 
\frac{1}{2 \, \alpha_{em}} \, 
\left( - \gamma_0(0) + \delta_{LT}(0) \right) \nonumber \\
&=&\,\frac{Q^6}{48 M_N^2} \frac{g_{A}^2}{(4\pi F_{\pi})^{2} \, m_{\pi}^2}
\left\{ -1 \,+\, {{m_\pi} \over {M_N}} \cdot {\pi \over 8} 
\left[ (27 + 7 \kappa_v) + (6 + 6 \kappa_s) \tau_3 \right] \right\} \, ,
\label{eq:g23limit} 
\end{eqnarray}
where $\gamma_0(0)$ is the forward spin polarizability 
at the real photon point, which has also been
calculated in HBChPT to ${\cal O}(p^{4})$, see Ref.~\cite{KSV02}. 
\newline
\indent
We have also studied the contribution of the $\Delta$(1232) resonance
in the ``small scale expansion'' ${\cal O}(\epsilon^{3})$, which uses
the quantity $\Delta = M_\Delta - M_N$ as a further expansion
parameter.  
Our results for the quantities $I_{1}$, $I_{A}$ and $\delta_{LT}$ are~:
\begin{eqnarray}
I_{1}^{\Delta}(Q^2)
&=&0, \nonumber \\
I_{A}^{\Delta}(Q^2)
&=&\frac{-Q^{2}}{18}(\frac{G_{1}}{\Delta})^{2} \nonumber \\
&-&\frac{8Q^{2}M^{2}}{9}\frac{g_{\pi\Delta N}^{2}}{(4\pi F_{\pi})^{2}}
\int^{1}_{0}dx \frac{x^2(1-2x)}{m_{0}^{2}}(\mu_{0}^{2}-1)^{-1}\left[1-
\mu_{0}\frac{\ln[\mu_{0}+\sqrt{\mu_{0}^{2}-1}]}{\sqrt{\mu_{0}^{2}-1}}
\right]\}, \\
\delta_{LT}^{\Delta}(Q^2)&=&
\frac{-32\alpha_{em}}{27}\frac{g_{\pi\Delta N}^{2}}{(4\pi F_{\pi})^{2}}
\int^{1}_{0}dx \;\frac{x^3}{m_{0}^{2}}(\mu_{0}^{2}-1)^{-2}\left[\mu_{0}^{2}
+2-3\mu_{0}\frac{\ln[\mu_{0}+\sqrt{\mu_{0}^{2}-1}]}{\sqrt{\mu_{0}^{2}-1}}
\right] \nonumber \\
&+&\frac{16\alpha_{em}}{9}\frac{g_{\pi\Delta N}^{2}}{(4\pi F_{\pi})^{2}}
\int^{1}_{0}dx \;\frac{x^{2}(1-2x)}{m_{0}^{2}}(\mu_{0}^{2}-1)^{-1}
\left[1-\mu_{0}\frac{\ln[\mu_{0}+\sqrt{\mu_{0}^{2}-1}]}{\sqrt{\mu_{0}^{2}-1}}
\right]\, , 
\label{eq:pol2delta}
\end{eqnarray}
with $m_{0}\equiv \sqrt{m_{\pi}^{2}+x(1-x)Q^2}$ 
and $\mu_{0}\equiv \frac{\Delta}{m_{0}}$.
Furthermore, 
$G_{1}$ and $g_{\pi\Delta N}$ are the leading order 
$\gamma \Delta N$ and $\pi\Delta N$ coupling constants respectively.
In the large $N_{C}$ limit of QCD, they are related 
with $\kappa_{v}$ and $g_{A}$ as~:
\begin{equation}
G_{1}=\frac{3}{2\sqrt{2}}\kappa_{v},
\,\,\,\,\,\,g_{\pi\Delta N}=\frac{3}{2\sqrt{2}}g_{A}.
\end{equation}
Furthermore, 
the ${\cal O}(\epsilon^3)$ $\Delta$ contribution to $d_2$ is
\begin{eqnarray}
d_2^{\Delta}(Q^2) &=& \frac{Q^6}{8 M_N^2} 
\left\{\frac{1}{18}(\frac{G_{1}}{M_{N}})^{2}\cdot\frac{1}{\Delta^2} 
\right. \nonumber \\
&&\hspace{1cm}-\frac{32}{27}\frac{g_{\pi\Delta N}^{2}}{(4\pi F_{\pi})^{2}}
\int^{1}_{0}dx \;\frac{x^3}{m_{0}^{2}}(\mu_{0}^{2}-1)^{-2}\left[\mu_{0}^{2}
+2-3\mu_{0}\frac{\ln[\mu_{0}+\sqrt{\mu_{0}^{2}-1}]}{\sqrt{\mu_{0}^{2}-1}}
\right] \nonumber \\
&&\left. \hspace{1cm} 
+\frac{24}{9}\frac{g_{\pi\Delta N}^{2}}{(4\pi F_{\pi})^{2}}
\int^{1}_{0}dx \;\frac{x^2(1-2x)}{m_{0}^{2}}(\mu_{0}^{2}-1)^{-1}\left[1-
\mu_{0}\frac{\ln[\mu_{0}+\sqrt{\mu_{0}^{2}-1}]}{\sqrt{\mu_{0}^{2}-1}}
\right]\right\}, \nonumber \\
\end{eqnarray}
and the ${\cal O}(\epsilon^3)$ $\Delta$ contribution to 
$\Gamma_1^{(3) \, inel}$ is
\begin{eqnarray}
\Gamma_1^{(3) \, \Delta}(Q^2) &=& \frac{Q^6}{8 M_N^2} 
\left\{\frac{-1}{18}(\frac{G_{1}}{M_{N}})^{2}\cdot\frac{1}{\Delta^2} 
\right. \nonumber \\
&&\left. \hspace{1cm}
-\frac{16}{27}\frac{g_{\pi\Delta N}^{2}}{(4\pi F_{\pi})^{2}}
\int^{1}_{0}dx \;\frac{x^3}{m_{0}^{2}}(\mu_{0}^{2}-1)^{-2}\left[\mu_{0}^{2}
+2-3\mu_{0}\frac{\ln[\mu_{0}+\sqrt{\mu_{0}^{2}-1}]}{\sqrt{\mu_{0}^{2}-1}}\right]\right\}. \nonumber \\
\end{eqnarray}
In the limit $Q^2 \to 0$, these quantities approache the values 
\begin{eqnarray}
d_2^{\Delta}(Q^2) &\longrightarrow& \frac{Q^6}{8 M_N^2} 
\left\{\frac{1}{16}(\frac{\kappa_{v}}{M_{N}})^{2}\cdot\frac{1}{\Delta^2} 
\right. \nonumber \\
&&\hspace{1cm}-\frac{1}{3}\frac{g_{A}^{2}}{(4\pi F_{\pi})^{2}m_{\pi}^{2}}
(\mu^{2}-1)^{-2}\left[\mu^{2}
+2-3\mu\frac{\ln[\mu+\sqrt{\mu^{2}-1}]}{\sqrt{\mu^{2}-1}}
\right] \nonumber \\
&&\left. \hspace{1cm} -\frac{1}{2}\frac{g_{A}^{2}}{(4\pi
F_{\pi})^{2}m_{\pi}^{2}}(\mu^{2}-1)^{-1}
\left[1-\mu\frac{\ln[\mu+\sqrt{\mu^{2}-1}]}{\sqrt{\mu^{2}-1}}\right]
\right\}, \\
\Gamma_1^{(3) \, \Delta}(Q^2) &\longrightarrow& \frac{Q^6}{8 M_N^2} 
\left\{\frac{-1}{16}(\frac{\kappa_{v}}{M_{N}})^{2}\cdot\frac{1}{\Delta^2} 
\right. \nonumber \\
&&\left. \hspace{1cm}
-\frac{1}{6}\frac{g_{A}^{2}}{(4\pi F_{\pi})^{2}m_{\pi}^{2}}
(\mu^{2}-1)^{-2}\left[\mu^{2}
+2-3\mu\frac{\ln[\mu+\sqrt{\mu^{2}-1}]}{\sqrt{\mu^{2}-1}}
\right]\right\} \, ,
\end{eqnarray}
with $\mu=\frac{\Delta}{m_{\pi}}$.
Analogous formulae can be derived for $\Gamma_2^{(3) \, \Delta}(Q^2)$ 
by taking the combination of Eq.~(\ref{eq:mom3g2equiv}).

\section{Multipole content of the moments}
\label{sec4}

By use of Eqs.~(\ref{eq:stt},\ref{eq:slt}), we can express the moment
$d_2$ by the partial cross sections $\sigma_{TT}$ and $\sigma_{LT}$~:
\begin{equation}
\label{eq:d2abs}
d_2^{inel} (Q^2) = \frac{Q^6}{32\pi^2\alpha_{em}M^2_N} \int_{\nu_0}^{\infty}
d\nu\,\frac{K(\nu,Q^2)}{\nu^2(\nu^2+Q^2)}\,
\left( -\sigma_{TT}(\nu,Q^2) +
\frac{3\nu^2+2Q^2}{\nu Q}\,\sigma_{LT}(\nu,Q^2) \right)\ ,
\end{equation}
and the multipole content follows by inserting the expansion
of these cross sections in terms of the electric (E), magnetic (M), 
and Coulomb or scalar (S) multipoles~\cite{DKT01}
\begin{eqnarray}
K\sigma_{TT} (1 \pi) & = & x_{TT}
 \sum_l \frac{1}{2}(l+1) \left[(l+2)(|E_{l+}|^2 +
|M_{l+1,-}|^2)\right.
\\ && \left.- l(|E_{l+1,-}|^2 + |M_{l+}|^2) + 2l(l+2)
(E_{l+}^{\ast}M_{l+}-
     E^{\ast}_{l+1,-}M_{l+1,-})\right]  \nonumber \\
& = & x_{TT}\,\left (|E_{0+}|^2 +|M_{1-}|^2 - |M_{1+}|^2 +
6E_{1+}^{\ast}M_{1+} + 3|E_{1+}|^2 \pm \ ... \right )\nonumber \ ,
\end{eqnarray}
\begin{eqnarray}
K\sigma_{LT} (1 \pi) & = & x_{LT}
\sum_l\frac{1}{2}(l+1)^2\\
&& \cdot \left[S_{l+}^{\ast}((l+2)E_{l+} + lM_{l+}) +
S_{l+1,-}^{\ast}
   (lE_{l+1,-} - (l+2)M_{l+1,-})\right] \nonumber
\\ & = & x_{LT}\,\left (S_{0+}^{\ast}E_{0+} - S_{1-}^{\ast}M_{1-}
+ 2S_{1+}^{\ast}(M_{1+}+3E_{1+})
  \pm\ ... \right )\nonumber \ ,
\end{eqnarray}
where
\begin{equation}
x_{TT} = x_{TT}(\nu,Q^2) = 4\pi\sqrt{(\nu-\nu_0)\,(\nu-\nu_0+2m_{\pi})}\ ,
\end{equation}
\begin{equation}
x_{LT} = x_{LT}(\nu,Q^2) = \sqrt{\frac{1+2\nu/M_N - Q^2/M_N^2}{1+\nu^2/Q^2}}
\,x_{TT}(\nu,Q^2)\ .
\end{equation}

The multipoles are generally functions of the $cm$ energy of the photon,
${\mathcal{M}}_{l\pm}\,(\omega,Q^2)$, with
\begin{equation}
\omega = \frac{M_N\nu-Q^2}{\sqrt{2M_N\nu + M_N^2 - Q^2}}\ .
\end{equation}

In comparing with Ref.~\cite{DKT01} it should be noted that we have changed
the sign of the partial cross sections in order to stay in accordance
with the usual notation of deep inelastic scattering, i.e.,
$\sigma_{TT} = -\sigma'_{TT}$ and $\sigma_{LT} = -\sigma'_{LT}$, 
where $\sigma'_{TT}$ and $\sigma'_{LT}$ were used in Ref.~\cite{DKT01}.  
We also note that expressions like $S_{0+}^{\ast}E_{0+}$ should be
read as ${\rm{Re}}\,(S_{0+}^{\ast}E_{0+})$.

The expression for the moments $\Gamma_1^{(3)}$, $\Gamma_2^{(3)}$ 
and $f_2$ can be derived similarly as Eq.~(\ref{eq:d2abs}).
\begin{equation}
\label{eq:g13abs}
\Gamma_1^{(3) \, inel} (Q^2) \,=\, 
\frac{Q^6}{32\pi^2\alpha_{em}M^2_N} \int_{\nu_0}^{\infty}
d\nu\,\frac{K(\nu,Q^2)}{\nu^2(\nu^2+Q^2)}\,
\left(\sigma_{TT}(\nu,Q^2) + \frac{Q}{\nu}\sigma_{LT}(\nu,Q^2) \right)\ ,
\end{equation}
\begin{equation}
\label{eq:g23abs}
\Gamma_2^{(3) \, inel} (Q^2) \,=\, 
\frac{Q^6}{32\pi^2\alpha_{em}M^2_N} \int_{\nu_0}^{\infty}
d\nu\,\frac{K(\nu,Q^2)}{\nu^2(\nu^2+Q^2)}\,
\left(- \sigma_{TT}(\nu,Q^2) + \frac{\nu}{Q}\sigma_{LT}(\nu,Q^2) \right)\ ,
\end{equation}
\begin{eqnarray}
\label{eq:f2abs}
f_2^{inel} (Q^2) & = & 
- \frac{9Q^2}{4M_N^2}\,\Gamma^{(1)}_{1, \, tw-2} \,+\,
\frac{9Q^4}{32\pi^2\alpha_{em}M^2_N}
\int_{\nu_0}^{\infty}
d\nu\,\frac{K(\nu,Q^2)} {\nu^2+Q^2}\,
\left( \sigma_{TT}(\nu,Q^2) + \frac{Q}{\nu}\sigma_{LT}(\nu,Q^2)
\right) \nonumber \\
&+& \frac{Q^6}{64\pi^2\alpha_{em}M^2_N} \int_{\nu_0}^{\infty}
d\nu\,\frac{K(\nu,Q^2)}{\nu^2(\nu^2+Q^2)} 
\left( \sigma_{TT}(\nu,Q^2) - \frac{3(\nu^2+Q^2)}{Q\nu}\sigma_{LT}(\nu,Q^2)
\right)\ .
\end{eqnarray}

\section{Results and discussion}
\label{sec5}

\begin{figure}[h]
\includegraphics[height=15cm]{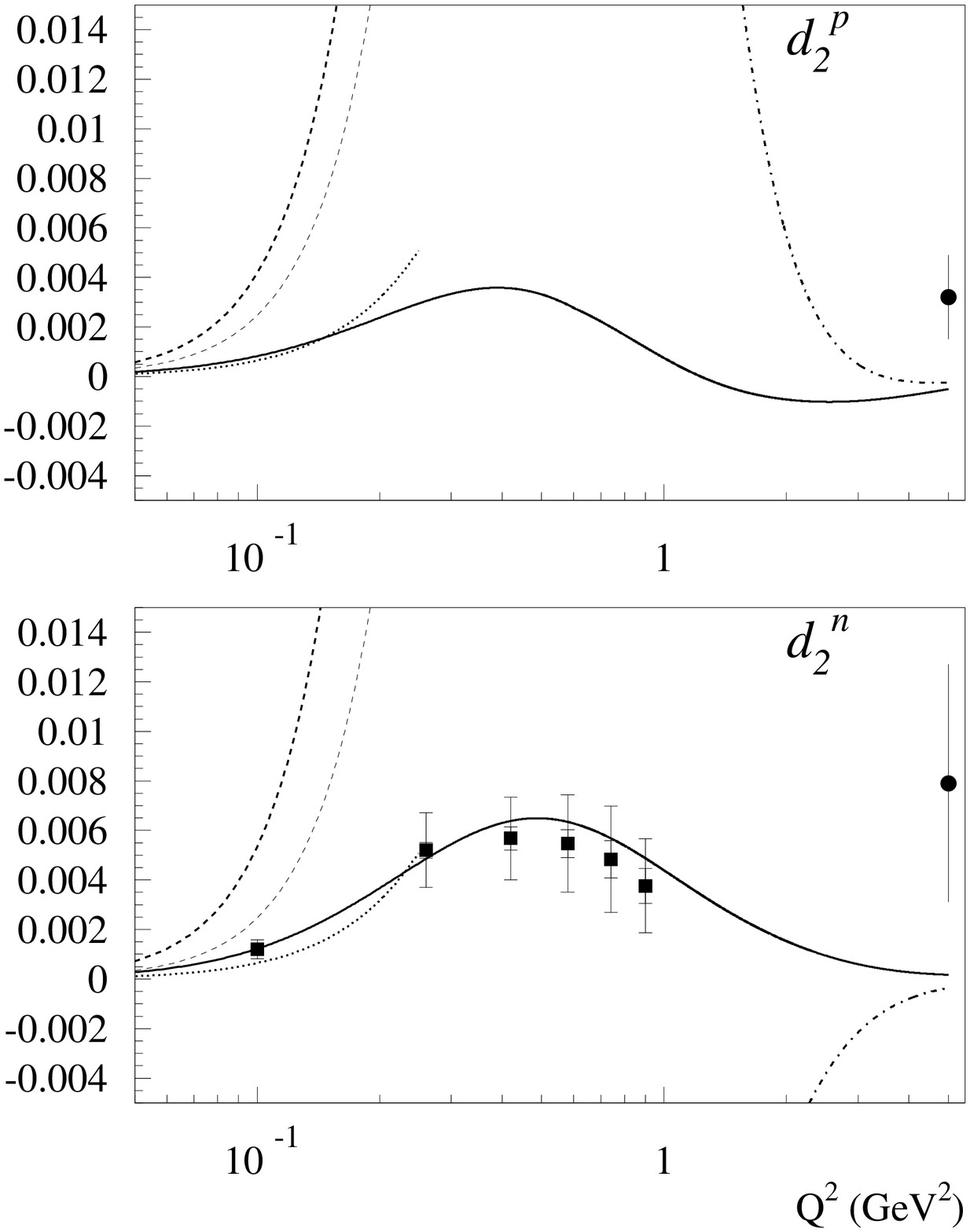}
\vspace{0cm}
\caption{$Q^2$ dependence of the moment $d_2$ 
for proton (upper panel) and neutron (lower panel).
The dashed-dotted curve is the elastic contribution to $d_2$ 
according to Eq.~(\ref{eq:d2el}). 
The other curves represent the inelastic contributions to $d_2$. 
Solid curves : MAID estimate for the $\pi$ channel; 
dotted curves : ${\mathcal O}(p^3)$ HBChPT; 
thick (upper) dashed curves : ${\mathcal O}(p^3) + {\mathcal O}(p^4)$ HBChPT;  
thin (lower) dashed  curves : ${\mathcal O}(p^3)$ HBChPT with 
${\mathcal O}(\varepsilon^3)$ $\Delta$ contribution added.  
The JLab/Hall A data (diamonds) are from Ref.~\protect\cite{Ama03} 
(inner error bars are statistical errors only, outer error bars include
systematical errors). The SLAC data (circles at $Q^2 = 5$~GeV$^2$) are 
from Ref.~\cite{E155X}.}  
\label{fig:d2}
\end{figure}

\begin{figure}[h]
\includegraphics[height=15cm]{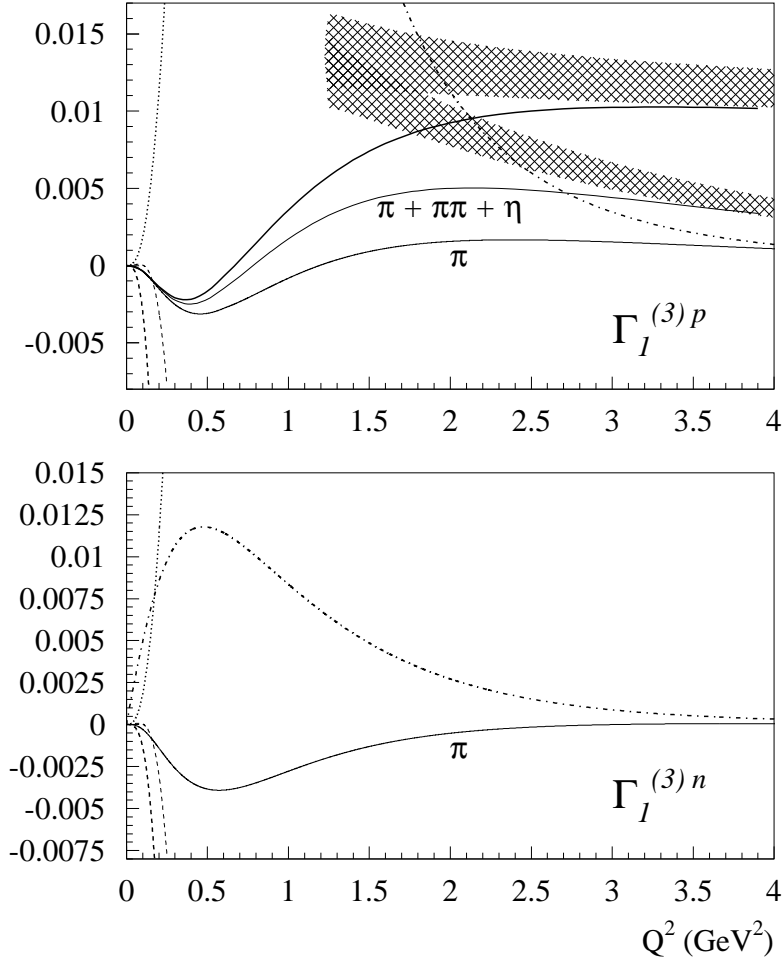}
\vspace{0cm}
\caption{$Q^2$ dependence of the moment $\Gamma^{(3)}_1$ 
for proton (upper panel) and neutron (lower panel).
The dashed-dotted curve is the elastic contribution to $\Gamma^{(3)}_1$  
according to Eq.~(\ref{eq:momg1el}). 
The other curves represent the inelastic contributions to $\Gamma^{(3)}_1$. 
Thin solid curves : MAID estimates for the $\pi$ and 
$\pi + \pi \pi + \eta$ channels (for proton) as indicated on curves; 
upper thick solid curve (for proton) : total estimate according to 
Eq.~(\protect\ref{eq:interpol}); 
dotted curves : ${\mathcal O}(p^3)$ HBChPT; 
thick dashed curves : ${\mathcal O}(p^3) + {\mathcal O}(p^4)$ HBChPT;
thin dashed  curves : ${\mathcal O}(p^3)$ HBChPT with 
${\mathcal O}(\varepsilon^3)$ $\Delta$ contribution added.  
The upper shaded band  is the 
evaluation using the DIS structure function $g_1$
as extracted from experiment, according to the analyis of 
Ref.~\protect\cite{Blum02}. 
The lower shaded band is the corresponding analysis 
for the contribution from the region $W < 2$~GeV (i.e. resonance region) 
to the DIS parametrization. 
The size of the bands represents the corresponding ($1 \sigma$) 
error estimates as given by Ref.~\protect\cite{Blum02}. 
}
\label{fig:mom3g1}
\end{figure}

\begin{figure}[h]
\includegraphics[height=15cm]{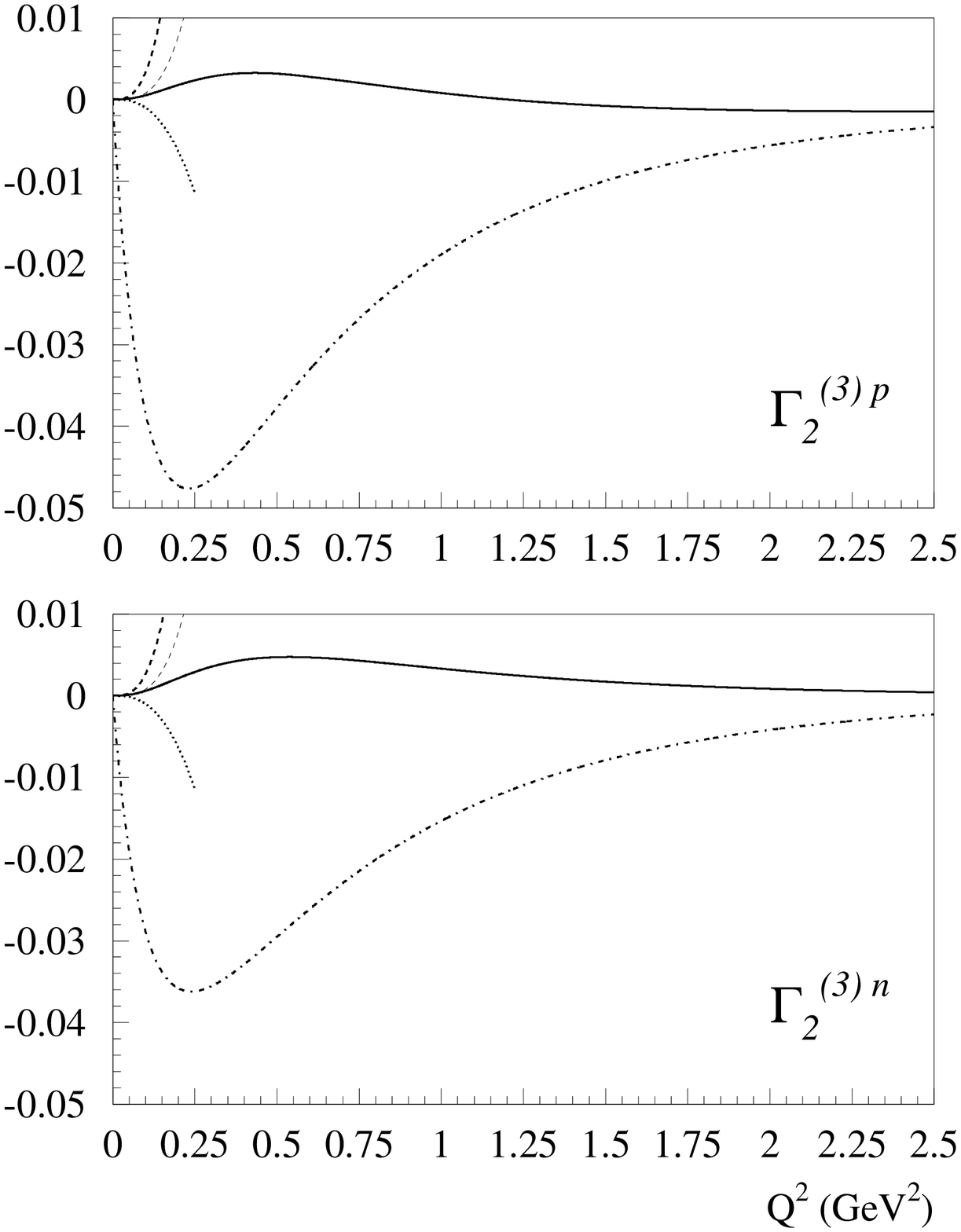}
\vspace{0cm}
\caption{$Q^2$ dependence of the moment $\Gamma^{(3)}_2$ 
for proton (upper panel) and neutron (lower panel).
The dashed-dotted curves are the elastic contributions to $\Gamma^{(3)}_2$  
according to Eq.~(\ref{eq:momg2el}). 
The other curves represent the inelastic contributions to $\Gamma^{(3)}_2$. 
Solid curves : MAID estimates for the $\pi$ channel; 
dotted curves : ${\mathcal O}(p^3)$ HBChPT; 
thick dashed curves : ${\mathcal O}(p^3) + {\mathcal O}(p^4)$ HBChPT;
thin dashed  curves : ${\mathcal O}(p^3)$ HBChPT with 
${\mathcal O}(\varepsilon^3)$ $\Delta$ contribution added.  
}
\label{fig:mom3g2}
\end{figure}

\begin{figure}[h]
\includegraphics[height=15cm]{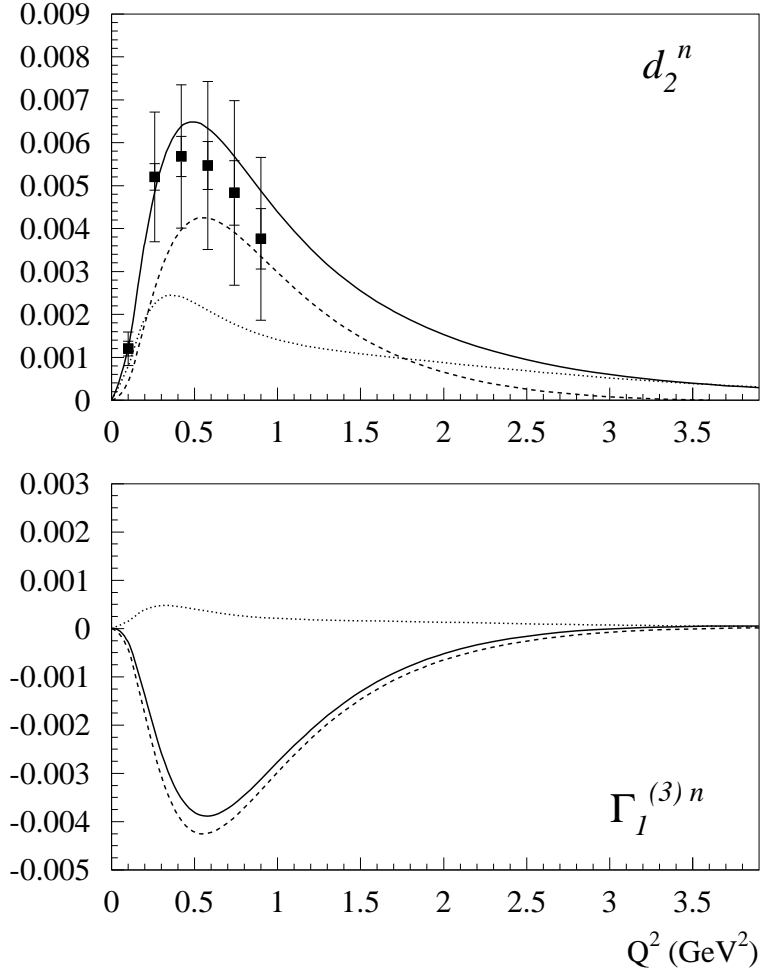}
\vspace{0cm}
\caption{Different photon absorption cross section contributions  
to $d_2$ (upper panel) 
and  $\Gamma^{(3)}_1$  (lower panel) for the neutron. 
The solid curves are the total 
MAID estimate for the $\pi$ channel. 
The dashed (dotted) curves are the contributions from  
$\sigma_{TT}$ ($\sigma_{LT}$) separately in 
Eqs.~(\ref{eq:d2abs}) and (\ref{eq:g13abs}).
Data for $d_2$ as in Fig.~\ref{fig:d2}.}
\label{fig:d2lt}
\end{figure}

\begin{figure}[h]
\includegraphics[height=15cm]{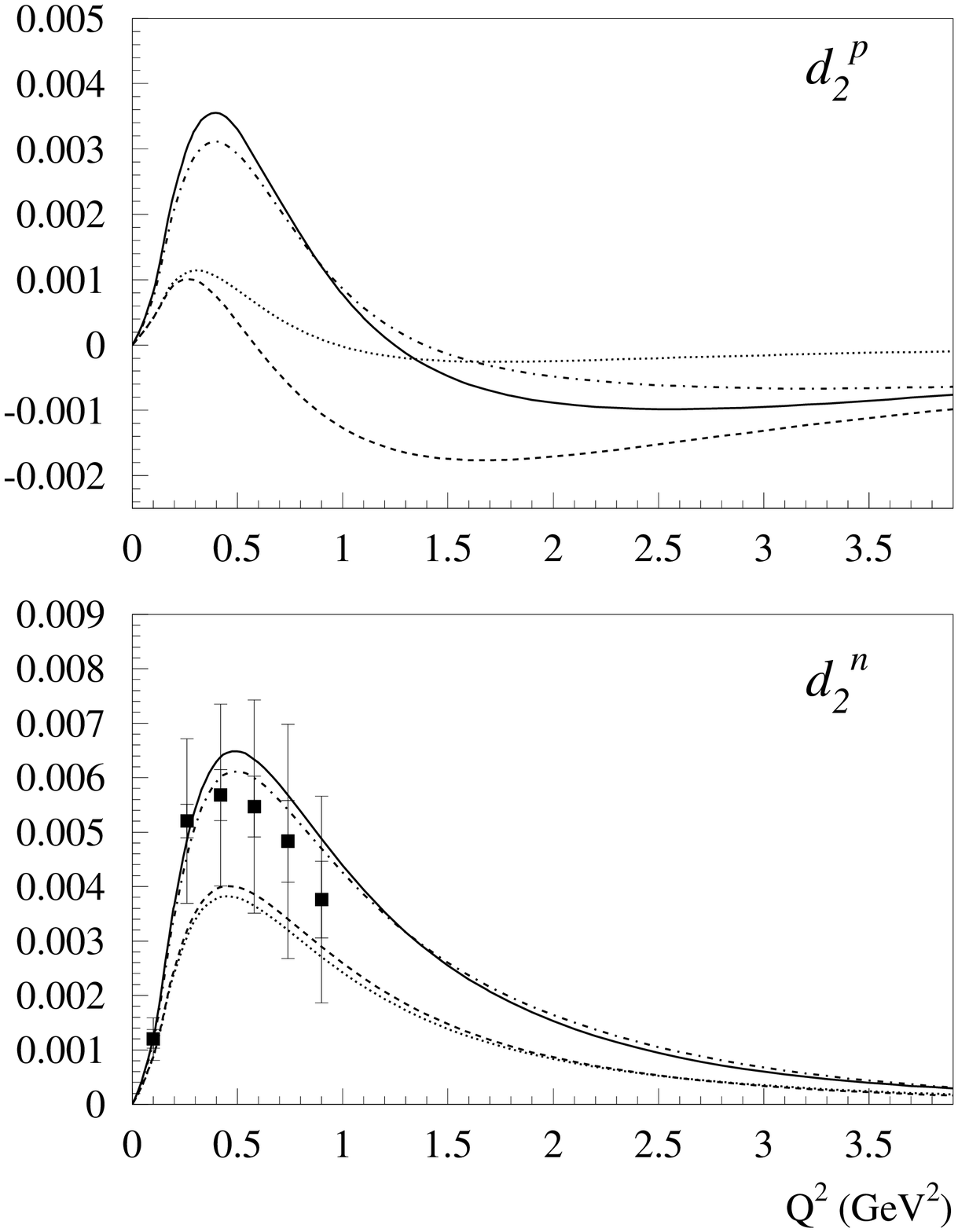}
\vspace{0cm}
\caption{Multipole content of the moment $d_2$ for proton (upper panel) 
and neutron (lower panel). 
The solid curves are the total 
MAID estimate for the $\pi$ channel (same as solid curves 
in Fig.~\ref{fig:d2}). 
The other curves represent different partial wave contributions. 
Dashed-dotted curves : results for the sum of s- and p-waves;
dashed curves : results for only s-waves; 
dotted curves : results for s-waves without $S_{11}(1535)$ resonance 
contribution. Data as in Fig.~\ref{fig:d2}.}
\label{fig:d2mult}
\end{figure}

\begin{figure}[h]
\includegraphics[height=15cm]{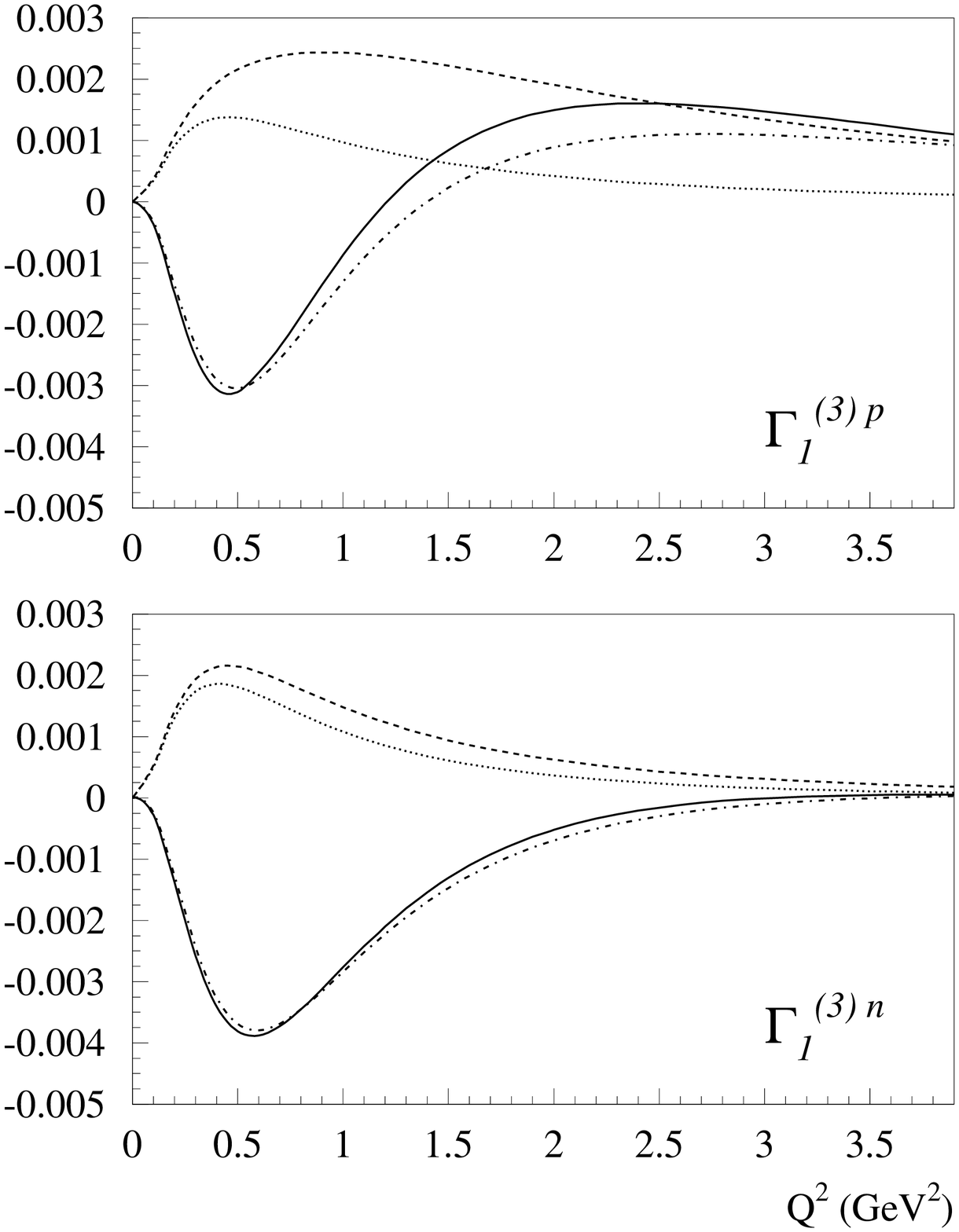}
\vspace{0cm}
\caption{Multipole content of the moment $\Gamma^{(3)}_1$  
for proton (upper panel) and neutron (lower panel). 
The solid curves are the total 
MAID estimates for the $\pi$ channel (same as 
corresponding solid curves in Fig.~\ref{fig:mom3g1}). 
The other curves represent different partial wave contributions. 
Dashed-dotted curves : results for the sum of s- and p-waves;
dashed curves : results for only s-waves; 
dotted curves : results for s-waves without $S_{11}(1535)$ resonance 
contribution.}
\label{fig:g13mult}
\end{figure}

\begin{figure}[h]
\includegraphics[height=15cm]{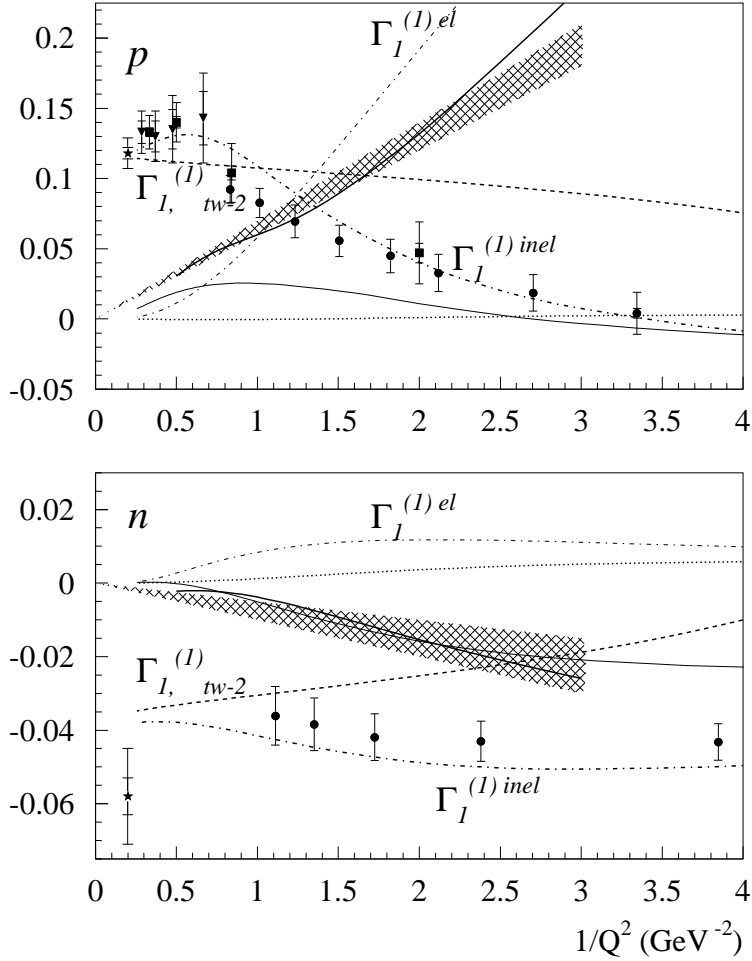}
\vspace{0cm}
\caption{$Q^2$ dependence of $f_2 \cdot 4 M_N^2 / (9 Q^2)$ 
(thick solid curves in the shaded bands) 
for proton (upper panel) and neutron (lower panel) as extracted from  
Eq.~(\ref{eq:f2def}). 
We show separately
the elastic $\Gamma^{(1) \, el}_1$  (thin dashed-dotted curves),  
total inelastic $\Gamma^{(1) \, inel}_1$ (thick dashed-dotted curves), 
and resonance ( $W < 2$ GeV ) (thin solid curves)   
contributions to $\Gamma^{(1)}_1$. 
We also show the  
twist-2 part $\Gamma^{(1)}_{1, \, tw-2}$ (dashed curves) 
and $(a_2 + 4 d_2) \cdot M_N^2 / (9 Q^2)$ (dotted curves) which enter 
on the {\it rhs} of Eq.~(\ref{eq:g1twist}) for $\Gamma^{(1)}_1$. 
The shaded bands are a linear fit to extract $f_2$ as 
described in the text. 
The proton data for $\Gamma_1^{(1) \, inel}$ are from 
JLab/CLAS \protect\cite{CLAS02} (circles), 
SLAC \protect\cite{Abe97} (diamonds), and 
HERMES \protect\cite{Air02} (triangles). 
The neutron data for $\Gamma_1^{(1) \, inel}$ are from 
JLab/HallA \protect\cite{Ama02} (circles). 
The data points at $Q^2 = 5$ GeV$^2$ (stars) correspond with the 
global analysis of Ref.~\cite{E155}. 
Inner error bars are statistical errors only, 
outer error bars include systematical errors. 
}
\label{fig:f2}
\end{figure}

In Fig.~\ref{fig:d2}, we show the results for the $Q^2$ dependence 
of the moment $d_2$.   
At large $Q^2$, the values of $d_2$ have been obtained for both
proton and neutron by DIS experiments at SLAC using a transversely
polarized target \cite{E155X}. From these experiments, the following 
values have been extracted at 
an average value of $Q^2 = 5$~GeV$^2$ \cite{E155X}~:
\begin{eqnarray}
d_2^p(Q^2 \approx 5 \, {\mathrm{GeV}}^2) \,&=&\, 0.0032 \pm 0.0017 \, , 
\nonumber\\
d_2^n(Q^2 \approx 5 \, {\mathrm{GeV}}^2) \,&=&\, 0.0079 \pm 0.0048 \, .
\end{eqnarray}
In the low and intermediate $Q^2$ region, we show the results for 
$d_2$ according to the generalized definition of Eq.~(\ref{eq:d2}). 
It rises strongly at the lower $Q^2$, displaying a $Q^6$ dependence 
at low $Q^2$, and tending to a small constant value asymptotically, 
corresponding with the twist-3 matrix element entering in the OPE 
of Eq.~(\ref{eq:g1twist}).   
Due to this structure, 
$d_2$ is an interesting observable to interpolate between 
a hadronic description at low $Q^2$ 
and a partonic description at large $Q^2$ values. 
One sees from Fig.~\ref{fig:d2} that   
the ChPT results rise strongly with $Q^2$, 
and its $Q^6$ dependence at the low $Q^2$ values is given by  
Eq.~(\ref{eq:d2limit}). The large difference between the 
$O(p^3)$ and $O(p^4)$ results originates from the known large difference 
between the $O(p^3)$ and $O(p^4)$ 
HBChPT results for the forward spin polarizability $\gamma_0$ 
\cite{JiKO0,Kum00,Gel00}. 
It is interesting to note however that the 
$O(p^3)$ HBChPT result is in good agreement with the 
phenomenological MAID estimate up to $Q^2 \simeq 0.25$~GeV$^2$. 
The convergence of the heavy baryon chiral expansion for the 
forward spin polarizability has been 
recently investigated in Ref.~\cite{BHM03} using the Lorentz 
invariant formulation of baryon ChPT of Ref.~\cite{BL99}. 
It was found~\cite{BHM03} that the main reason for the 
slow convergence of $\gamma_0$ in the $1/M_N$ expansion is due to 
the slow convergence of the Born graphs. The corresponding 
one-loop relativistic calculation cures this deficiency. 
However the relativistic result to fourth order in the chiral expansion, 
even when supplemented with $\Delta$ and vector meson contributions,  
still does not agree with the data, suggesting to envisage a  
fifth-order calculation for $\gamma_0$ in future work.  
\newline
\indent 
Very recently, the inelastic contribution to $d_2$ has been measured for the 
neutron at intermediate $Q^2$ values at JLab/Hall A \cite{Ama03}. 
It is seen from Fig.~\ref{fig:d2} that the phenomenological resonance 
prediction using the MAID model for the neutron shows  
an excellent agreement with these data. We will discuss the MAID results  
for $d_2^n$ further on, to get some better understanding of the origin of 
this good description. 
\newline
\indent
In Figs.~\ref{fig:mom3g1} and \ref{fig:mom3g2}, we show the results for the 
$Q^2$ dependence of the third moments of the nucleon spin structure 
functions $g_1$ and $g_2$ respectively. 
From Eqs.~(\ref{eq:a2limit},\ref{eq:g23limit}), 
we see that the inelastic part to 
the moment $\Gamma_1^{(3)}$ is proportional to $Q^6 \, \cdot \, \gamma_0$ 
at small $Q^2$, and that the inelastic part to 
the moment $\Gamma_2^{(3)}$ also contains $\gamma_0$ at small $Q^2$.     
Therefore the HBChPT results for those moments 
directly reflect the poor convergence of 
the chiral expansion for $\gamma_0$. The MAID model  
predicts a value for $\gamma_0$ at the real photon point as~:
$\gamma_0^p = -0.707 \cdot 10^{-4} \mathrm{fm}^4$, which  
is in relative good agreement with 
the experimental value (for its determination, 
see Ref.~\cite{DPV03})~: 
\begin{equation}
\label{DDeq2.2.19}
\gamma_0^p = \left[ -1.01 \pm 0.08~(\mathrm{stat}) \pm 0.10~(\mathrm{syst})
\right] \cdot10^{-4}~\mathrm{fm}^4 \ .
\end{equation}
Going to larger values of $Q^2$, the MAID results for the proton 
change sign. At large $Q^2$, the value of $\Gamma_1^{(3)}$ is known 
from DIS experiments as shown on Fig.~\ref{fig:mom3g1}. We also compare 
in Fig.~\ref{fig:mom3g1} the MAID result with the resonance contribution 
to $\Gamma_1^{(3)}$ (corresponding with $W <2$~GeV) 
as extracted from the fit of $g_1$ data. 
From this comparison, we see that the 
$\pi$-channel alone underestimates the total resonance contribution at larger 
$Q^2$. For the proton channel, we are also able to provide an estimate 
for the $\eta N$ and $\pi \pi N$ channels which provide the dominant 
virtual photon absorption cross sections beyond $\pi N$. 
It is seen that the 
sum of the $\pi + \pi \pi + \eta$ channels nicely joins the 
$W < 2$~GeV part of the DIS parametrization (lower shaded band 
in Fig.~\ref{fig:mom3g1}) for $Q^2 > 3$~GeV$^2$.  
To gain an understanding of the gradual transition in $\Gamma_1^{(3)}$, 
from the resonance dominated to the partonic regime, 
we generalize a suggestion of Ref.~\cite{Ans89} for the first moment $I_1$ of 
$g_1$ to parametrize its $Q^2$ dependence 
through a vector meson dominance type model. 
This model for $I_1$ was refined in 
Refs.~\cite{Bur92,Bur93} by adding explicit resonance contributions 
which are important at low $Q^2$ as seen above.  
We can generalize this phenomenological parametrization 
for the inelastic part to an arbitrary moment 
of $g_1$ through the interpolating formula~:
\begin{eqnarray}
\Gamma_1^{(n) \, inel}(Q^2) \,&=& \, \Gamma_1^{(n) \, res}(Q^2) \,+\, 
\frac{(Q^2)^n}{\left( Q^2 + \mu^2 \right)^{n + 1}} \, 
\left( Q^2 \, \Gamma_1^{(n) \, as} \, +\, \mu^2 \, c^{(n)} \right) \, ,
\label{eq:interpol}
\end{eqnarray}
where $\Gamma_1^{(n) \, res}(Q^2)$ is the resonance contribution evaluated 
using the MAID model,  
and $\mu$ is a vector meson mass scale ($\mu \approx 0.77$ GeV), which 
governs the transition to the asymptotic value $\Gamma_1^{(n) \, as}$. 
The coefficient $c^{(n)}$ is chosen in such a way that $\Gamma_1^{(n)}$ 
reaches its known value at the real photon point. 
For $c^{(1)}$, this is fixed by the non-resonant contribution to the 
GDH sum rule as~:  
\begin{eqnarray}
c^{(1)} \,=\, \frac{\mu^2}{2 \, M_N^2} \, \left( \, -\kappa^2/4 \,-\,
I_1^{res}(0) \, \right),
\label{eq:c1}
\end{eqnarray}
where $I_1^{res}(0)$ is the resonance contribution ($W < 2$~GeV) 
at the real photon point to $I_1$.
For the third moment, the coefficient $c^{(3)}$ is fixed by the 
very small non-resonant contribution to the forward spin polarizability~:
\begin{eqnarray}
c^{(3)} \,=\, \frac{(\mu^2)^3}{16 \, M_N^2} \, \frac{1}{\alpha_{em}} \, 
\left( \, \gamma_0 (0) \,-\, \gamma_0^{res} (0) \, \right) ,
\end{eqnarray}
where $\gamma_0^{res}(0)$ is the resonance contribution (corresponding with 
$W < 2$~GeV).
Analogously, one can express the coefficients $c^{(n)}$ for 
higher moments ($n = 5, 7,...$) through the very small non-resonant 
contributions to the higher spin polarizabilities of the nucleon. 
As these higher polarizabilities are practically completely dominated 
by the low-energy excitation region (i.e. resonance region), 
it is a very good approximation to take $c_1^{(n)} \simeq 0$ for 
$n \geq 5$.
\newline
\indent
In Fig.~\ref{fig:mom3g1}, we also show the result of  
the interpolating model of Eq.~(\ref{eq:interpol}) for 
$\Gamma_1^{(3)}$ of the proton, 
by using the MAID model for the resonance 
contribution $\Gamma_1^{(3) \, res}$ and using as asymptotic 
contribution for the proton~: $\Gamma_1^{(3) \, as} \simeq 0.012$. 
It will be interesting to compare these predictions with experimental 
results which can be extracted from corresponding analyses for the 
lowest moments of the proton \cite{CLAS02} and neutron \cite{Ama02} 
structure function $g_1$.
\newline
\indent
Besides the inelastic contributions, 
we also show in Figs.~\ref{fig:d2},\ref{fig:mom3g1} 
for comparison the elastic contributions to $d_2$ and $\Gamma_1^{(3)}$ 
according to Eqs.~(\ref{eq:d2el},\ref{eq:momg1el}). 
The elastic contribution is calculated by using for the proton the 
form factor parametrization of
Ref.~\cite{Bra02} including the new JLab data for the ratio of 
$G_E^p / G_M^p$~\cite{Jon00,Gay02}.
For the neutron, the elastic contribution is calculated using  
for $G_E^n$ the recent fit of Ref.~\cite{War02} (following the Galster form), 
and for $G_M^n$ the recent fit of Ref.~\cite{Kub02}. 
We see from Fig.~\ref{fig:d2} that for the proton the elastic 
contribution largely dominates the 
inelastic one for both $d_2$ and $\Gamma_1^{(3)}$ when $Q^2 < 1$~GeV$^2$. 
For the neutron, 
the elastic contributions to $d_2$ and $\Gamma_1^{(3)}$ are of comparable size 
to the inelastic ones and vanish at the real photon point. 
\newline
\indent
In Fig.~\ref{fig:d2lt}, we show the separate contributions of 
$\sigma_{TT}$ and $\sigma_{LT}$
to $d_2^n$ and $\Gamma_1^{(3)n}$ to gain some insight in the nature of the 
photon absorption mechanism involved. 
In the case of the twist-3 moment $d_2$ of Eq.~(\ref{eq:d2abs}), 
$\sigma_{LT}$ is enhanced by the kinematical term in front, 
and therefore becomes dominant at large values of $\nu$ and $Q^2$. 
The situation is quite different
for the third moment $\Gamma_1^{(3)n}$ of Eq.~(\ref{eq:g13abs}), 
in which case the LT term vanishes 
at $Q^2=0$ and is suppressed for large $\nu$.
\newline
\indent
The multipole content of $d_2$ is displayed in Fig.~\ref{fig:d2mult} 
for both proton and neutron.
This moment is dominated by an interplay of s- and p-wave multipoles up to 
$Q^2\approx4$~GeV$^2$. The striking difference between the proton and 
neutron is essentially due to the excitation of the $S_{11}(1535)$ resonance. 
According to the
Particle Data Group~\cite{Hag02}, 
the helicity amplitudes for the $S_{11}(1535)$ are
$A_{1/2}^p = (0.090 \pm 0.030)$~GeV$^{-1/2}$ and 
$A_{1/2}^n = (-0.046 \pm 0.027)$~GeV$^{-1/2}$, 
with the result that the cross sections
for $S_{11}$ excitation on the proton are about a factor 4 larger 
as compared to the neutron. 
It is also known that this resonance
has a very hard form factor \cite{Arm99,Tho01}, 
and therefore this effect persists up to 
very high momentum transfers. 
This leads to the difference that $d_2^p$ changes sign around 
$Q^2\approx 1.25$~GeV$^2$, whereas $d_2^n$ remains at the same sign as 
confirmed by the data. 
\newline
\indent
Fig.~\ref{fig:g13mult} shows the multipole 
analysis for the moment  $\Gamma_1^{(3)n}$. The
conclusions are similar as in the previous case. The strong coupling of the
$S_{11}$ to the proton leads to the distinct feature that $\Gamma_1^{(3)p}$
changes sign at $Q^2\approx 1.25$~GeV$^2$ and asymptotically remains at larger
values than for the neutron.
\newline
\indent
In Fig.~\ref{fig:f2}, we show our result for $f_2$ as extracted from 
Eq.~(\ref{eq:f2def}). To calculate $f_2$, we need the 
full $Q^2$ dependence of $\Gamma^{(1)}_1$ consisting of a sum of 
elastic and inelastic contributions. The elastic contribution 
is given by Eq.~(\ref{eq:momg1el}) and evaluated using the most recent 
experimental information on the nucleon form factors as detailed above. 
The inelastic contribution can e.g. be obtained by using the  
interpolating formula of Eq.~(\ref{eq:interpol}). Although this gives a 
decent description of both proton and neutron data, 
a better fit for the proton data is obtained by 
using as interpolating formula~:
\begin{eqnarray}
\Gamma_1^{(1) \, inel}(Q^2) \,&=& \, \Gamma_1^{(1) \, res}(Q^2) \,+\, 
\Gamma_1^{(1) \, as} \, \tanh \left( \frac{Q^2 \, (Q^2 + \mu^2 \, c^{(1)} / \Gamma_1^{(1) \, as})}{\mu^2 \, (Q^2 + \mu^2)} \right) \, ,
\label{eq:interpol2}
\end{eqnarray}
where $c^{(1)}$ is as defined in Eq.~(\ref{eq:c1}), with  
$I_1^{res, \, p}(0) = -0.67$ obtained from the MAID model.  
Furthermore, we fix the asymptotic value by using 
$\Gamma_1^{(1) \, p} = 0.118$ at $Q^2 = 5$~GeV$^2$, 
as obtained from the global analysis of Ref.~\cite{E155}. 
This value then fixes the asymptotic coefficient 
$\Gamma_1^{(1) \, as}$ in the interpolating formula for 
$\Gamma_1^{(1) \, inel}$, as well as 
the flavor singlet axial charge $a_0^\infty$ in the 
twist-2 part $\Gamma_{1, \, tw-2}^{(1)}$ of Eq.~(\ref{eq:mu2}).
The shape parameter $\mu$ in 
Eq.~(\ref{eq:interpol2}) is taken as $\mu \simeq 0.74$~GeV, 
which gives a very good description of all proton data for 
$\Gamma_1^{(1) \, inel}$ as seen from Fig.~\ref{fig:f2}.  
Note that we merely consider the functional form of 
Eq.~(\ref{eq:interpol2}) as a convenient representation of the inelastic data. 
\newline
\indent
For the neutron, we use the interpolating formula of Eq.~(\ref{eq:interpol}), 
with $I_1^{res, \, n}(0) = -0.48$ obtained from the MAID model.  
Furthermore, we use as asymptotic value $\Gamma_1^{(1) \, as, n} = -0.038$, 
and mass scale $\mu \simeq 0.57$~GeV. Note that in order to get a good 
description of the JLab/Hall A data of Ref.~\cite{Ama02},  
we use an asymptotic value which is somewhat 
smaller in absolute value than the value 
$\Gamma_1^{(1) \, n} (Q^2 = 5~\mathrm{GeV}^2) 
= -0.058 \pm 0.005 \pm 0.008$ 
as obtained from the global analysis of Ref.~\cite{E155}. 
\newline
\indent
Furthermore in Fig.~\ref{fig:f2}, we also show separately the twist-2 part 
$\Gamma^{(1)}_{1, \, tw-2}$, and 
the term $(a_2 + 4 d_2) \cdot M_N^2 / (9 Q^2)$, which is seen to be 
negligibly small for both proton and neutron. Therefore $f_2$ is
dominated by the difference 
$\Gamma^{(1) \, el}_1 + \Gamma^{(1) \, inel}_1 - \Gamma^{(1)}_{1, \, tw-2}$.  
The quantity $f_2 \cdot 4 M_N^2 / (9 Q^2)$ 
evaluated using Eq.~(\ref{eq:f2def}) 
is shown by the thick solid curves in Fig.~\ref{fig:f2}. 
We can then extract $f_2$ from the linear regime in the $1/Q^2$ plot. 
Because the slight non-linear structure observed in the thick solid curves 
in Fig.~\ref{fig:f2} is within the error bars of the data, we 
make a linear fit by including all data for $Q^2$ in the range 
$0.5 - 2$~GeV$^2$, in particular the new JLab data for both proton and 
neutron.  
This is shown by the shaded bands in the same figure, which 
fully accomodate the thick solid curves in the mentioned $Q^2$ range. 
From the slopes of these bands, we can then extract $f_2$ 
for $Q^2$ in the range 0.5 - 2 GeV$^2$ as~:  
\begin{eqnarray}
f_2^p &\,\simeq\,& 0.15 \, \to \, 0.18 \, , \nonumber \\
f_2^n &\,\simeq\,& -0.026 \to -0.013\, . 
\label{eq:f2values}
\end{eqnarray}  
\indent
The twist-4 matrix elements $f_2$ have been extracted before in 
Ref.~\cite{JiMel97} from the data of Ref.~\cite{Abe96} at $Q^2$ = 1 GeV$^2$, 
resulting in the values
~\footnote{Note that in the present work, we follow the opposite sign 
convention for $f_2$ as in Ref.~\cite{JiMel97}.}~: 
\beqn
f_2^p & = & 0.10\pm0.05\ , \nonumber \\
f_2^n & = & 0.07\pm0.08\ .
\label{eq:f2valuesmel}
\eeqn
Comparing our results of Eq.~(\ref{eq:f2values}) with those of 
Ref.~\cite{JiMel97}, 
we see that they are compatible for the proton 
but that we extract a small negative value for the neutron, 
whereas in Ref.~\cite{JiMel97} a positive result was extracted. 
However, given the uncertainty of the neutron data both values of 
$f_2^n$ might well be consistent with zero.  
One sees from Fig.~\ref{fig:f2}, that for the neutron there is a 
partial cancellation between the elastic and inelastic contribution to 
$\Gamma_1^{(1)}$, and the result for $f_2$ is practically completely given 
by the resonance contribution $\Gamma_1^{(1) \, res}$, which is estimated here 
in the MAID model and gives a good description of the available neutron data 
as we have seen before. This might partly explain why we extract a 
negative value for $f_2^n$, compared with the analysis of Ref.~\cite{JiMel97}. 
\newline
\indent
We can compare these phenomenological values for $f_2$ with several model 
estimates. In Ref.~\cite{Ste95}, an estimate was given within QCD sum rules 
which yielded as values~: 
$f_2^p = -0.037 \pm 0.006$ 
and $f_2^n = -0.013 \pm 0.006$.  
In Ref.~\cite{Lee02}, the instanton vacuum picture was used to estimate matrix 
elements of quark gluon operators. In particular it has been shown in this 
approach that the matrix element $d_2$ is suppressed relative to $f_2$ as~:
\begin{equation}
\frac{d_2}{f_2} \sim \left( \frac{\bar \rho}{\bar R} \right)^4 << 1 \, ,
\end{equation}
where $\bar \rho / \bar R \simeq 1/3$ is the average instanton density. 
Using the $SU(3)$ symmetric case, the twist-4 matrix elements $f_2$ were 
obtained in Ref.~\cite{Lee02} as~: $f_2^p = -0.046$ 
and $f_2^n = +0.038$.  We notice that in the approach of Ref.~\cite{Lee02}, 
proton and neutron values are of comparable size and opposite sign. 
In both the QCD sum rules and instanton vacuum models, the proton value 
is smaller and of opposite sign compared with the 
phenomenologically extracted value of Ref.~\cite{JiMel97} and with the value 
Eq.~(\ref{eq:f2values}) of this work. 
\newline
\indent
It has been suggested that the 
twist-3 moment $d_2$ and the twist-four moment $f_2$ are related to  
the response of the color electric ($\chi_E$) and magnetic ($\chi_B$)
fields to the polarization of the nucleon in its 
rest frame~\cite{Ji95,Man95}, defined as~:
\beqn
\langle P S | \psi^\dagger \, g \vec B \, \psi | P S \rangle \,&=&\, 
\chi_B \, 2 M_N^2 \, \vec S \, , \nonumber \\
\langle P S | \psi^\dagger \, \vec \alpha \times g \vec E \, \psi | P S \rangle \,&=&\, 
\chi_E \, 2 M_N^2 \, \vec S \, ,
\eeqn
where $P$ is the nucleon momentum and $S$ the 
projection of its spin vector $\vec S$. Furthermore $\vec E$ ($\vec B$) 
are the color electric (magnetic) fields respectively, and $g$ is 
the strong coupling constant.  
Furthermore, the moments $d_2$ and $f_2$ 
can be expressed in terms of the gluon-field polarizabilities as~:
\beqn
d_2 & = & \frac{1}{4}\,(\chi_E + 2\chi_B)\ , \nonumber \\
f_2 & = & (\chi_E - \chi_B)\ .
\label{polglue}
\eeqn
Since the experimental value of $d_2$ is of order $10^{-3}$, 
we see from Eq.~(\ref{eq:f2values}) that the predicted central values 
of $f_2$ are larger than those of $d_2$ by about a factor 50 for the proton. 
These findings agree, at least qualitatively, with estimates using QCD sum
rules~\cite{Ste95} and based on the instanton vacuum approach~\cite{Lee02}, 
but less so with the predictions of bag models~\cite{JiUn94}. 
Within the large uncertainties of all existing predictions, 
we may therefore conclude that $d_2\ll f_2$. 
This observation can be combined with Eq.~(\ref{polglue}) to yield
\beqn
\chi_E & \approx & + \frac{2}{3}\,f_2\ ,\nonumber \\
\chi_B & \approx & - \frac{1}{3} \,f_2\ .
\eeqn
In particular a positive value of $f_2$, 
such as found in both the phenomenological extraction of 
Eq.~(\ref{eq:f2values}) and Eq.~(\ref{eq:f2valuesmel}) for the proton, 
leads to a negative value of $\chi_B$, i.e., color diamagnetism.

\section{Conclusions}
\label{sec6}

In this work we have studied higher moments of nucleon spin 
structure functions in view of recent data at intermediate values of $Q^2$. 
In particular, we evaluated the generalizations 
to arbitrary momentum transfer $Q^2$ of 
the third moment $d_2$ of the twist-3 part of the nucleon spin 
structure function $g_2$, as well as the third moments 
$\Gamma_1^{(3)}$ and $\Gamma_2^{(3)}$ of 
$g_1$ and $g_2$ respectively.  
We have shown that the physical interpretation of 
$d_2$, $\Gamma_1^{(3)}$, and $\Gamma_2^{(3)}$  
at arbitrary values of $Q^2$ is given through  
nucleon generalized (i.e. $Q^2$ dependent) spin polarizabilities. 
These higher moments in the Bjorken variable $x$ 
contain appreciable contributions from the resonance region.
We therefore evaluated these moments in heavy baryon chiral perturbation theory
(HBChPT) at order ${\mathcal{O}}(p^4)$ and in a unitary isobar model (MAID).  
\newline
\indent
The ChPT results were found to rise strongly with $Q^2$, 
and display a $Q^6$ dependence at low $Q^2$ values, proportional to the 
forward spin polarizabilities at the real photon point.  
For $d_2$, the $O(p^3)$ HBChPT results are in good agreement with the 
phenomenological MAID estimate up to $Q^2 \simeq 0.25$~GeV$^2$. 
However there is a large difference between the 
$O(p^3)$ and $O(p^4)$ results. This difference originates 
from the known large difference between the $O(p^3)$ and $O(p^4)$ 
HBChPT results for the forward spin polarizability $\gamma_0$, 
for which the chiral expansion is poorly converging. 
\newline
\indent
The phenomenological unitary isobar prediction using the MAID model 
shows an excellent agreement with recent neutron data from JLab/Hall A 
at intermediate $Q^2$ values. 
The good description is found in part 
to be due to a sizeable contribution from the 
$\sigma_{LT}$ photoabsorption cross section. 
Furthermore, we performed a multipole expansion of these higher 
moments to gain some additional insight in the dominant absorption mechanisms.
We found that both $d_2$ and $\Gamma_1^{(3)}$ are dominated 
by an interplay of s- and p-wave multipoles up to $Q^2\approx4$~GeV$^2$, 
and observed a striking difference between the proton and 
neutron, which is essentially due to the 
excitation of the $S_{11}(1535)$ resonance. 
The helicity amplitudes for the photoexcitation of the 
$S_{11}$ on the proton are about a factor 4 larger 
as compared to the neutron, and this resonance
has a very hard form factor. 
This leads to the result that both $d_2^p$ and $\Gamma_1^{(3) \, p}$ 
change sign around $Q^2\approx 1.25$~GeV$^2$, 
and asymptotically remain at much larger values 
than in the case of the neutron. 
It will be interesting to test this prediction by extracting 
$d_2$ and $\Gamma_1^{(3)}$ from proton data, 
and to see if one obtains a similar good description as for the neutron. 
\newline
\indent
Besides the twist-3 matrix element $d_2$, the $1/Q^2$ suppressed term in the 
twist expansion of the first moment $\Gamma_1^{(1)}$ of $g_1$ also contains  
the twist-4 matrix element $f_2$. 
By including all data for $\Gamma_1^{(1)}$ for $Q^2$ in the range 
$0.5 - 2$~GeV$^2$, in particular the new JLab data 
for both proton and neutron, we extracted $f_2$ in this $Q^2$ range  as~:  
$f_2^p \,:\, 0.15 \, \to \, 0.18$, and 
$f_2^n \,:\, -0.026 \to -0.013$. 
\newline
\indent
The values of $d_2$ and $f_2$ 
enter in the response of the color electric and 
magnetic fields to the polarization of the nucleon in its rest frame. 
Therefore our numerical estimates yield phenomenological predictions 
for gluon field polarizabilities, which are new nucleon structure information. 
\newline
\indent
In summary, the sum rules which we studied in this work as function of $Q^2$ 
from the real photon point to the region of deep inelastic scattering, 
are very interesting  
observables to interpolate between the non-perturbative 
and perturbative regimes of QCD. As these higher moments do not 
require large extrapolations into unmeasured regions, they are 
ideal observables to be measured at intermediate momentum transfers.

\section*{Acknowledgements}

This work was supported by the Deutsche Forschungsgemeinschaft (SFB443) 
and by the U.S. Department of Energy under contract 
DE-AC05-84ER40150. 
The authors like to thank G. Cates, J.-P. Chen, S. Choi, Z.-E. Meziani 
for useful discussions and correspondence.


\end{document}